\documentclass[twocolumn]{aastex62}
\usepackage{amsmath}
\usepackage{hyperref}
\definecolor{linkcolor}{rgb}{0.02,0.35,0.55}
\definecolor{citecolor}{rgb}{0.45,0.45,0.45}
\hypersetup{colorlinks=true,linkcolor=linkcolor,citecolor=citecolor, filecolor=linkcolor,urlcolor=linkcolor}
\hypersetup{pageanchor=true}
\newcommand{\beq}{\begin{equation}\begin{aligned}}
\newcommand{\eeq}{\end{aligned}\end{equation}}

\shorttitle{Spatial Line Offsets in PDRs: Carina, Cyg OB2 and IC 1396}
\shortauthors{Carlsten and Hartigan}

\begin{document}

\title{Photoevaporation of Molecular Clouds in Regions of Massive Star Formation as Revealed Through H$_2$ and Br$\gamma$ Emission}

\author{S.G. Carlsten}
\altaffiliation{scottgc@astro.princeton.edu}
\affiliation{Physics and Astronomy Department, Rice University, 6100 Main Street, Houston, TX, 77005, USA}
\affiliation{Department of Astrophysical Sciences, Princeton University, 4 Ivy Lane, Princeton,
  NJ 08544, USA}
\author{P.M. Hartigan}
\affiliation{Physics and Astronomy Department, Rice University, 6100 Main Street, Houston, TX, 77005, USA}

\begin{abstract}

We examine new and pre-existing wide-field, continuum-corrected, narrowband images in H$_2$ 1-0 S(1) and Br$\gamma$ of three regions of massive star formation: IC 1396, Cygnus OB2, and Carina. These regions contain a variety of globules, pillars, and sheets, so we can quantify how the spatial profiles of emission lines behave in photodissociation regions (PDRs) that differ in their radiation fields and geometries. We have measured 450 spatial profiles of H$_2$ and Br$\gamma$ along interfaces between HII regions and PDRs. Br$\gamma$ traces photoevaporative flows from the PDRs, and this emission declines more rapidly with distance as the radius of curvature of the interface decreases, in agreement with models. As noted previously, H$_2$ emission peaks deeper into the cloud relative to Br$\gamma$, where the molecular gas absorbs far-UV radiation from nearby O-stars.  Although PDRs in IC 1396, Cygnus OB2, and Carina experience orders of magnitude different levels of ionizing flux and have markedly differing geometries, all the PDRs have spatial offsets between Br$\gamma$ and H$_2$  on the order of $10^{17}$cm. There is a weak negative correlation between the offset size and the intensity of ionizing radiation and a positive correlation with the radius of curvature of the cloud.  We can reproduce both the size of the offsets and the dependencies of the offsets on these other variables with simple photoevaporative flow models. Both Br$\gamma$ and H$_2$ 1-0 S(1) will undoubtedly be targeted in future JWST observations of PDRs, so this work can serve as a guide to interpreting these images.

\end{abstract}


\section{Introduction} \label{sec:intro}

As stars coalesce out of dense concentrations within molecular clouds,
ultraviolet radiation, primarily from the most massive
young stars, slowly erodes the clouds away.
Between the ionized region that immediately surrounds young stars and the much colder
ambient molecular cloud, far UV (FUV) stellar radiation with energies between $\sim$ 6~eV and 13.6~eV
dominates the chemistry and heating of the gas.  These FUV photons dissociate molecular hydrogen 
in the cloud but do not have enough energy to ionize atomic hydrogen. The resulting
``photodissociation regions" (PDRs) where the cloud absorbs stellar ultraviolet radiation
expand in photoevaporative flows and produce a
wide-variety of pillar shapes and other complex geometries \citep[e.g.][]{hartigan2015}. \par

Newborn stars also inject momentum and energy into molecular clouds 
through protostellar outflows \citep[e.g.][]{quillen2005}, stellar winds \citep[e.g.][]{harper2009}, 
and radiation pressure \citep[e.g.][]{lopez2011,pellegrini2007, lopez2014, krumholz2009, 
krumholz2014}. Photoionization and winds are both negative feedback mechanisms in the sense that they 
energize and drive turbulence in the parent molecular cloud \citep{matzner2002, krumholz2006, krumholz2014}
and generally reduce the ability of the cloud to create new stars. However, in some cases the 
HII region may drive a shock wave into the cloud and compress the gas enough to
trigger new stars to form \citep{bertoldi89,sicilia2014}.
Overall, the feedback mechanisms of photoionization and stellar outflows
explain why star formation rates are much lower than expected from gravitational
collapse alone \citep[e.g.][]{zuckerman1974, krumholz2007, evans2009}. Feedback also
makes up a key component of modern N-body simulations of galaxy
formation \citep[e.g.][]{schaye2015,genel2014,vogelsberger2014,springel2003}. {For a review of feedback processes and their implementation in numerical simulations of star and galaxy formation, see \citet{dale2015}.} \par

In this work, we study the HII regions and PDRs present in three large regions of
massive star formation: Cygnus OB2, NGC 3372 (Carina Nebula) and IC 1396. These
three regions represent massive star formation on very different scales, 
allowing us to investigate how different radiation fields from young stars influence molecular clouds.
Cygnus OB2, part of the larger Cygnus X region, is one of the most active regions of massive star
formation in the Milky Way as measured by stellar content. Studies have catalogued at least 169
primary OB stars in Cygnus OB2 with a total stellar mass of $\sim$ $1.6\times 10^{4}$$M_{\odot}$
and {total molecular cloud mass of $\sim$ $7\times 10^5$M$_\odot$ \citep{wright2015,schneider2006,comeron2012}. Due to the high level of extinction, many more early-type stars are likely still undiscovered in the association \citep{berlanas2018}.} Based on its
size, some have even suggested considering Cygnus OB2 as a young galactic globular cluster instead of a
massive star association \citep{knod2000,comeron2002}. Star formation in the Cyg OB2 region appears to have started
$\sim$7 Myr ago and peaked 5 Myr ago \citep{wright2015}, though it
continues in the region today both within isolated globules and as evidenced by a scattered
group of over 1000 protostars \citep{kryukova2014}. { \citet{schneider2016} characterize the population of globules, pillars, and condensations in Cygnus OB2 and suggest an evolutionary path whereby pillars evolve into globules which can evolve into proplyd-like objects. We assume a distance to the Cygnus OB2 region of 1400pc \citep{rygl2012}. }\par

Carina is of a similar scale to Cygnus~OB2, with about
70 O stars catalogued \citep{smith2006}, including the famous LBV star $\eta$ Car. {The cloud mass of the Carina region is roughly $6\times10^{5}$M$_\odot$ with $>10^4$M$_\odot$ at the density sufficient to form more stars \citep{preibisch2012}}. Carina appears to be a
slightly younger 1-3 Myr old region \citep{hur2012} with copious ongoing star formation { that might be triggered by the massive star clusters \citep{smith2010, rocc2013, gac2013}.}
As such, Carina offers an example of a younger analog of Cygnus OB2. {In this paper, we focus on the surroundings of Trumpler 16 and Trumpler 14, the two most massive young clusters \citep{smith2006}. We assume a distance to Carina of 2300pc \citep{smith2006}. }\par

In contrast to Cygnus
OB2 and Carina, IC 1396 has only a few massive stars. Similar to the Orion Nebula,
the primary source of ionization in IC~1396 is a single
O6.5V star (in this case HD 206267, part of the Tr~37 cluster). IC 1396 notably includes 
a famous bright rimmed cloud known as the Elephant Trunk Nebula (IC 1396A),
and there is evidence for sequential
star formation in the region. \citet{getman2012} found a spatial age gradient whereby the youngest stars
are located near the cloud, including inside of it, and stellar age increases as one
moves east towards Tr~37. They argue that HD~206267 caused a radiative driven implosion
over the past several Myr that sequentially formed these stars. \citet{sicilia2014} found a very
young protostellar object directly behind the ionization front (IF) consistent with this picture.
Tr~37 itself is found to have an episodic star formation history and is estimated to have a mean age of 4 Myr \citep{sicilia2005, sicilia2015}. {We assume a distance to IC 1396 of 870pc \citep{contreras2002}.}

Two of the brightest emission lines in the near-infrared spectrum of PDR's, 
H$_{2}$ 1-0 S(1) and Br$\gamma$, have proved to be excellent tracers of PDR interface shapes
because they highlight where the FUV radiation is absorbed while penetrating through most of
the obscuring dust along the line of sight.  The 2.12$\mu$m H$_{2}$ emission line
is a fluorescent line of molecular hydrogen that is emitted in a cascade following the
absorption of a FUV photon.  Photons that have energy above roughly 10eV can excite an
electron in H$_{2}$ into the Lyman and Werner bands, and, as the electron cascades 
to the ground state, there is a chance it passes through the transition that emits
a 2.12$\mu$m photon.  Hence, this line traces exactly where FUV radiation is absorbed in the cloud.
In contrast, the 2.17$\mu$m Br$\gamma$ line is a
recombination line of ionized hydrogen and thus traces the H$^{+}$ in the regions. 
Because the lines are bright and differential extinction is negligible, 
once any K-band continuum is removed the resulting images in Br$\gamma$ and H$_2$ 
allow us to probe the physical structure of the ionization fronts and PDRs throughout
a region of star formation.  These emission lines will undoubtedly be primary choices
for this type of work once JWST is launched.

In this paper we present extensive maps of
continuum-corrected H$_{2}$ 1-0 S(1) and Br$\gamma$ line emission in Cygnus OB2, Carina, and
IC 1396. Motivated by the distinct spatial offset between the H$_{2}$ and Br$\gamma$
emission observed in Carina's PDR interfaces \citep{hartigan2015}, we identify roughly 450 
locations along the PDR interfaces in the three star forming regions that are suitable for
extracting high signal-to-noise spatial profiles in both lines.
We then use numerical PDR models to understand how the size of the offset is determined by
the geometries and radiation fields present in these regions. 
In Section \ref{sec:data} we discuss the data acquisition
and reduction; in Section \ref{sec:Analysis}, we discuss the regions and the offsets we find in
each; and in Section \ref{sec:models} we introduce the numerical models and parameters used to compare with
and interpret our observations. Section \ref{sec:conclusions}
 summarizes our results and conclusions.  
 
\begin{deluxetable}{lcc}
\tablecaption{Summary of Observations\label{tab:observations}
}
\tablehead{
\colhead{Region} & \colhead{Dates} & \colhead{Telescope}
}
\startdata
Cyg OB2 & Sept 19-22, 2008 & 4m Mayall, KPNO\\
           & Nov 12-20, 2012 & 4m Mayall, KPNO\\
Carina     & March 11-18, 2011     & 4m Blanco, CTIO\\
IC 1396    & Oct 27-30, 2012     & 4m Mayall, KPNO\\
           & Nov 12-20, 2012     & 4m Mayall, KPNO\\
\enddata
\end{deluxetable}

\section{Data} \label{sec:data}

\begin{figure*}
\includegraphics[width=0.95\textwidth]{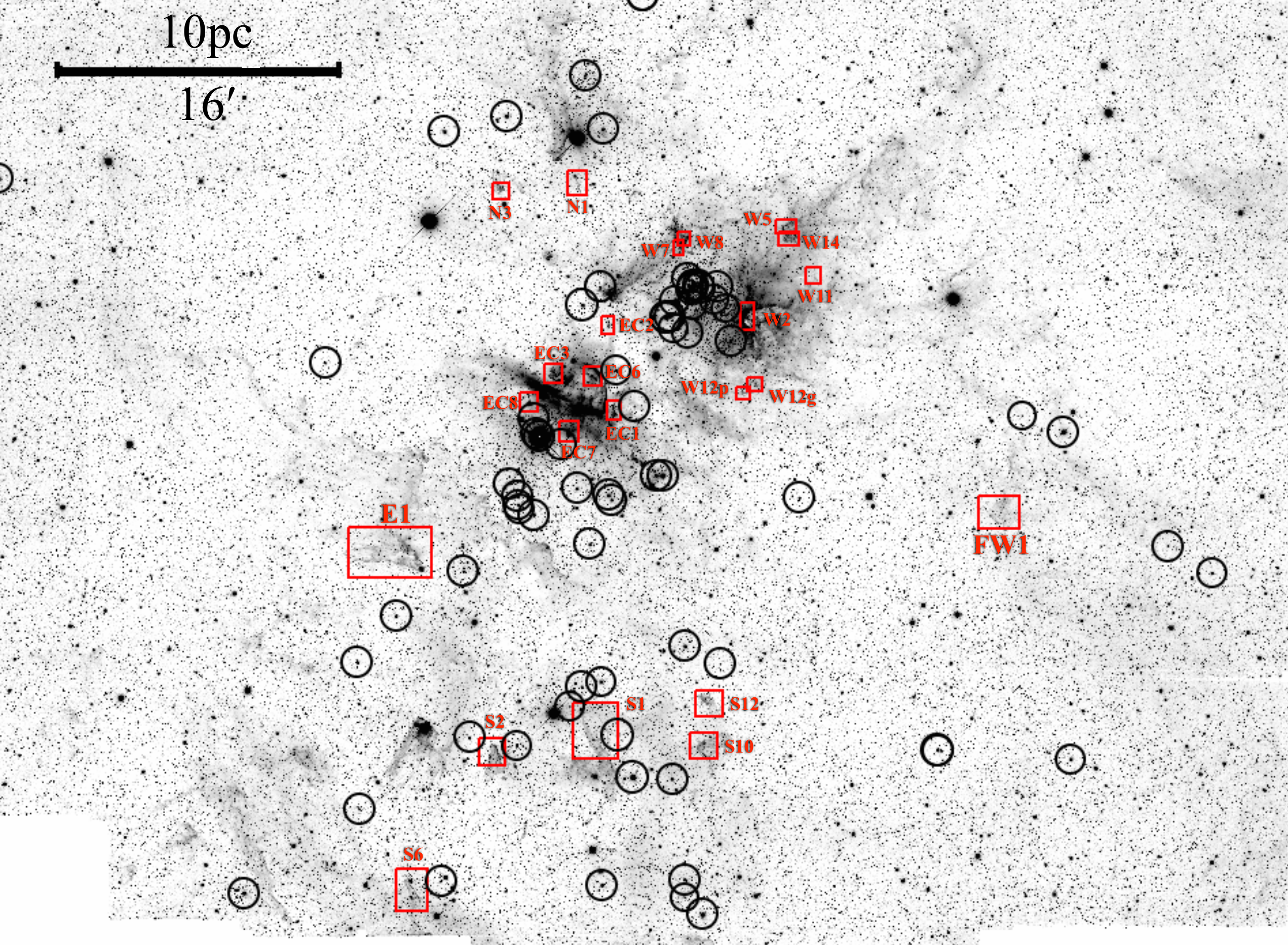}
\caption{H$_{2}$ mosaic of Carina region from \citet{hartigan2015}. Irradiated molecular interfaces appear as the dark pillars
and walls. The continuum has not been subtracted. The WR, O, and early
B stars from \citet{gagne2011} are shown in the black circles. {The molecular cloud interfaces analyzed in this work are indicated by the red boxes}. North is up and East is to the left in all the images. }
\label{fig:carina_img}
\end{figure*}

\subsection{Reduction}

The near infrared (NIR) data used in this project were all taken with the NOAO
NEWFIRM instrument. The imager consists of four 2048x2048 pixel detectors with a gap of 35
arcseconds between chips \citep{probst2008}. Attached to the NOAO 4m telescopes, the imager has a pixel
scale of 0.4 arcsec/pix. Details for the data acquisition for the three regions are given
in Table \ref{tab:observations}. NEWFIRM has a data
reduction pipeline \citep{swaters2009} that attempts a sky subtraction by using a median filter over the
nearest 3 images in time before and after a frame is taken. As discussed in
\citet{hartigan2015}, the pipeline is inadequate for regions of large-scale nebulosity.  We therefore re-reduce the data
in the manner outlined in \citet{hartigan2015}. To improve the S/N of our K-band mosaic, we used
an archival NEWFIRM image of IC 1396A taken in 2009 by T. Megeath.
This improved the S/N of the tip of IC 1396A significantly but did not help
for regions further down the trunk. The maps of Carina were presented and thoroughly discussed
in \citet{hartigan2015}, and several of the H$_2$ images of globules in Cygnus OB2 were shown
in \citet{hartigan2012}.

\begin{figure*}
\includegraphics[width=0.95\textwidth]{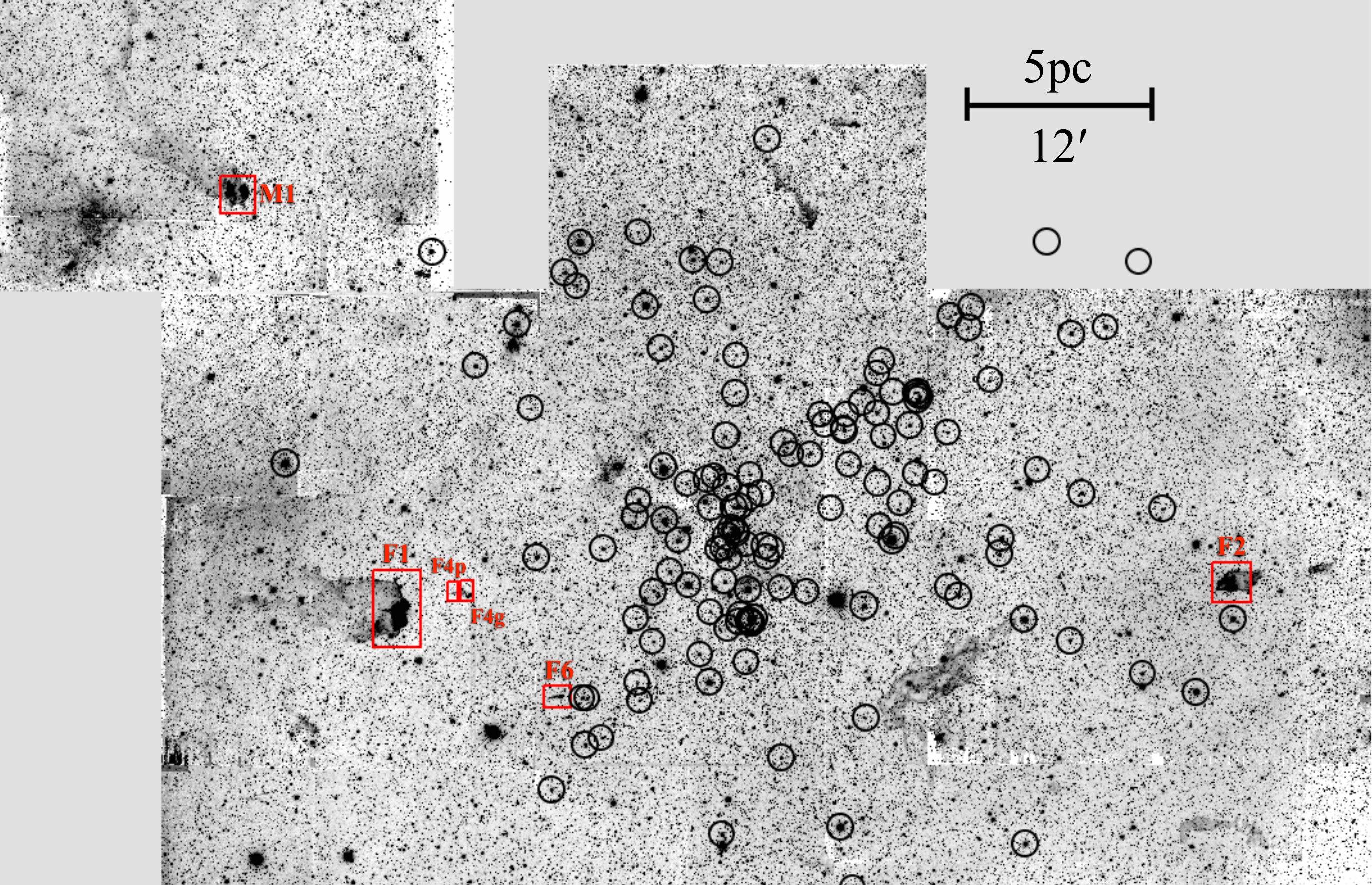}
\caption{H$_{2}$ mosaic of Cygnus OB2 region. The location of the WR, O, and early B
stars from \citet{wright2015} are shown in the black circles. {The molecular cloud interfaces analyzed in this work are indicated by the red boxes}. The continuum has not been subtracted.}
\label{fig:cyg_img}
\end{figure*}

The final stacked results were then aligned between the three filters. S/N and exposure time
varied within the frames due to the complicated dither patterns. The Carina data had average
seeing of 1.09", IC 1396 had 1.1", and Cyg OB2 had 1.1". 
The images were acquired through a variety of sky conditions and are not flux-calibrated, but
this aspect does not affect the spatial offsets observed between H$_{2}$ and Br$\gamma$ emission
or the spatial profile of either line, which are the focus of this paper.

\subsection{Continuum Subtraction}

The H$_{2}$ and Br$\gamma$ images are contaminated with continuum emission 
that arises primarily from starlight reflected from dust in the molecular clouds. 
We can remove the continuum by subtracting an appropriately scaled K-band image from
both the Br$\gamma$ and H$_2$ mosaics. Because the wavelengths of H$_2$ (2.12$\mu$m) and Br$\gamma$ (2.17$\mu$m)
are nearly identical and the continuum arises from the Rayleigh-Jeans part of the black-body
spectrum for a usual star, the amount of continuum is nearly proportional to the bandpasses of
each narrow-band filter. In this situation, there are three unknowns: the isolated
H$_{2}$ and Br$\gamma$ emission and the pure continuum emission, and three constraints: the total emission
recorded in each of the three filters. It is therefore possible to do filter
``algebra'' to isolate the line emission. In particular, we can write:

\begin{equation}
H_{2}^{tot} = H_{2} + \alpha K \\
\label{eq:h2}
\end{equation}

\begin{equation}
Br\gamma^{tot} = Br\gamma + \beta K \\
\label{eq:brg}
\end{equation}

\begin{equation}
K^{tot} = H_{2} + Br\gamma+  K \\
\label{eq:k}
\end{equation}

\noindent
where X$^{tot}$ is the observed flux in the X-filter image and X is the contribution of the
emission line. The constants $\alpha$ and $\beta$ are related through the bandpass widths of the H$_{2}$
and Br$\gamma$ filters, respectively. We determine their ratio empirically by constructing
scaled subtractions between
the H$_{2}^{tot}$ and Br$\gamma^{tot}$ images. Stars will generally not have any line emission in H$_{2}$
or Br$\gamma$ and will subtract out with the correct values of $\alpha$ and $\beta$.
We find that H$_{2}^{tot}$ - $C\times$Br$\gamma^{tot}$, where $C\equiv\alpha/\beta$, optimally removes stars when $C$=1.19 for all regions in
Carina and Cygnus except for one grid in Cygnus where $C$=1.40 is required. For IC 1396,
$C$ was found to be between 0.85 and 1.05 depending on the region of the
mosaic. The scaling factor includes the effects of different
filter bandpasses, exposure times, and sensor sensitivities. 

For some constant, $D$, adding $D$ times Eq. \ref{eq:brg} to Eq. \ref{eq:h2} and subtracting $D$
times Eq. \ref{eq:k} yields:

\begin{equation}
H_{2}^{tot} + DBr\gamma^{tot} - DK^{tot}= H_{2}(1-D) + K(\alpha + \frac{\alpha}{C}D-D)
\label{eq:deq}
\end{equation}

When $D$ is chosen such that the stars are properly subtracted, we see that the
coefficient of $K$ on the right hand side of Eq. \ref{eq:deq} must be zero, implying:

\begin{equation}
\alpha = \frac{D}{\frac{D}{C}+1}
\end{equation}

$\beta$ can then be found from the measured values of $C$. Once we know $\alpha$ and $\beta$, the line emission is given by:

\begin{equation}
H_{2} = \frac{H_{2}^{tot}-\frac{\alpha}{1-\beta}\times(K^{tot}-Br\gamma^{tot})}{1-\frac{\alpha}{1-\beta}}
\end{equation}

\begin{equation}
K = \frac{K^{tot} - H_{2} - Br\gamma^{tot}}{1-\beta}
\end{equation}

\begin{equation}
Br\gamma = Br\gamma^{tot} - \beta K \\
\end{equation}

$\alpha$ and $\beta$ are slightly different for the different mosaics but always close to the
ratios in bandpass between the H$_{2}$/Br$\gamma$ filters and the K filter. Our $\alpha$ and
$\beta$ for Carina correspond very closely to those found by \citet{yeh2015} who imaged 30 Dor
with NEWFIRM. 

\begin{figure}
\includegraphics[width=0.45\textwidth]{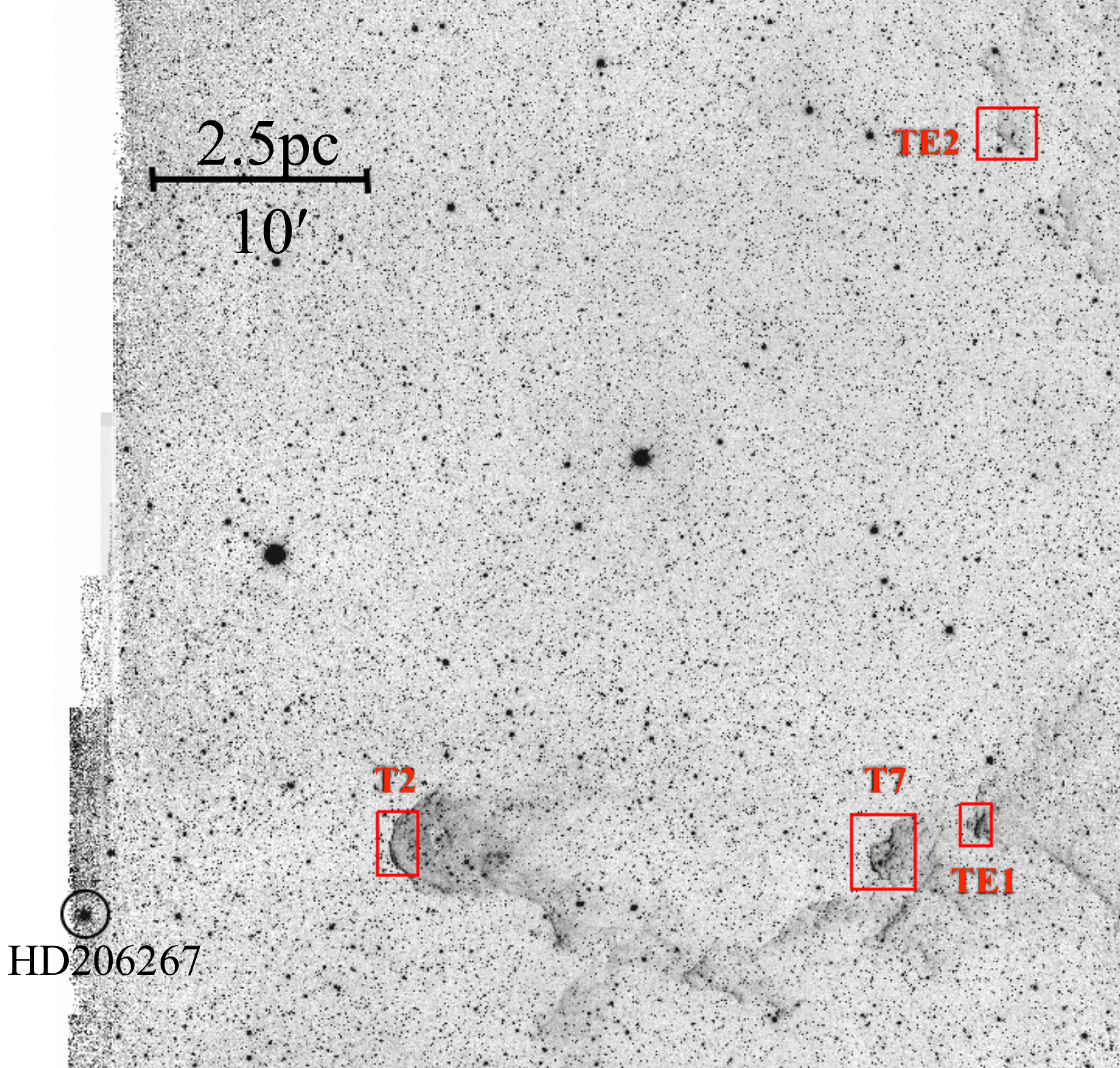}
\caption{H$_{2}$ mosaic of IC 1396 region. Location of HD206267 in the center of Trumpler 37
is denoted by a black circle. IC 1396A, the Elephant Trunk nebula, is visible across the bottom of the frame pointing towards HD206267. Part of the edge of the cavity is visible in the upper right. {The molecular cloud interfaces analyzed in this work are indicated by the red boxes.} The continuum has not been subtracted.}
\label{fig:ic_img}
\end{figure}

\section{Analysis} \label{sec:Analysis}
{Our primary goal in this paper is to quantify how ionization fronts (IFs) and
photodissociation regions (PDRs) appear in Carina, Cyg OB2, and IC 1396.
In this section we summarize the general characteristics of the H$_2$ emission-line
nebulae in each region (Section \ref{sec:h2_overview}), investigate how the spatial
profiles of Br$\gamma$ emission behave near the cloud surfaces (Section \ref{sec:brgProfiles}) and
consider the observed spatial offsets between H$_2$ and Br$\gamma$ across the molecular
cloud surfaces (Section \ref{sec:offsets}). Our study quantifies differences in the size of this
offset between the three star forming regions as well as trends in offset size
with cloud geometry and irradiation environment.}

\subsection{The PDR Interfaces in Carina, Cyg OB2, and IC 1396}\label{sec:h2_overview}

{The H$_2$ images are excellent tracers of PDR interfaces, and reveal
a variety of pillars, walls, and globules in Carina, Cyg OB2, and
IC 1396 (Figs. \ref{fig:carina_img}-\ref{fig:ic_img}). Each star-forming region has its own general
character in the H$_2$ images. The image of Carina
in Figure \ref{fig:carina_img} contains several clusters, including Tr 14, the bright
LBV star $\eta$ Car, and a wide variety of pillars, walls, and globules.
In contrast, though Cyg~OB2 also has large-scale H$_2$ walls, the brightest
PDRs are associated with irradiated globules located in the vicinity of the
concentration of massive stars in the middle of the images.}

{The last region, IC 1396A, is dominated by the Elephant Trunk
Nebula, a long pillar structure.} Also present are the edges of a large, circular bubble surrounding Tr 37. 
Because of the low S/N of
the K-band image, we focus primarily on the PDR interfaces along and at the base
of the trunk where the S/N is the best in the K-band. In all of our regions 
the Br$\gamma$ emission is located closer to the ionizing sources than the H$_{2}$,
compatible with the picture of molecular clouds being ionized and dissociated by radiation from the
newborn stars. For many of the bright interfaces, streams of photoionized gas are visible
in Br$\gamma$ as they evaporate from the surface of the clouds. 

\begin{figure*}
\includegraphics[width=0.99\textwidth]{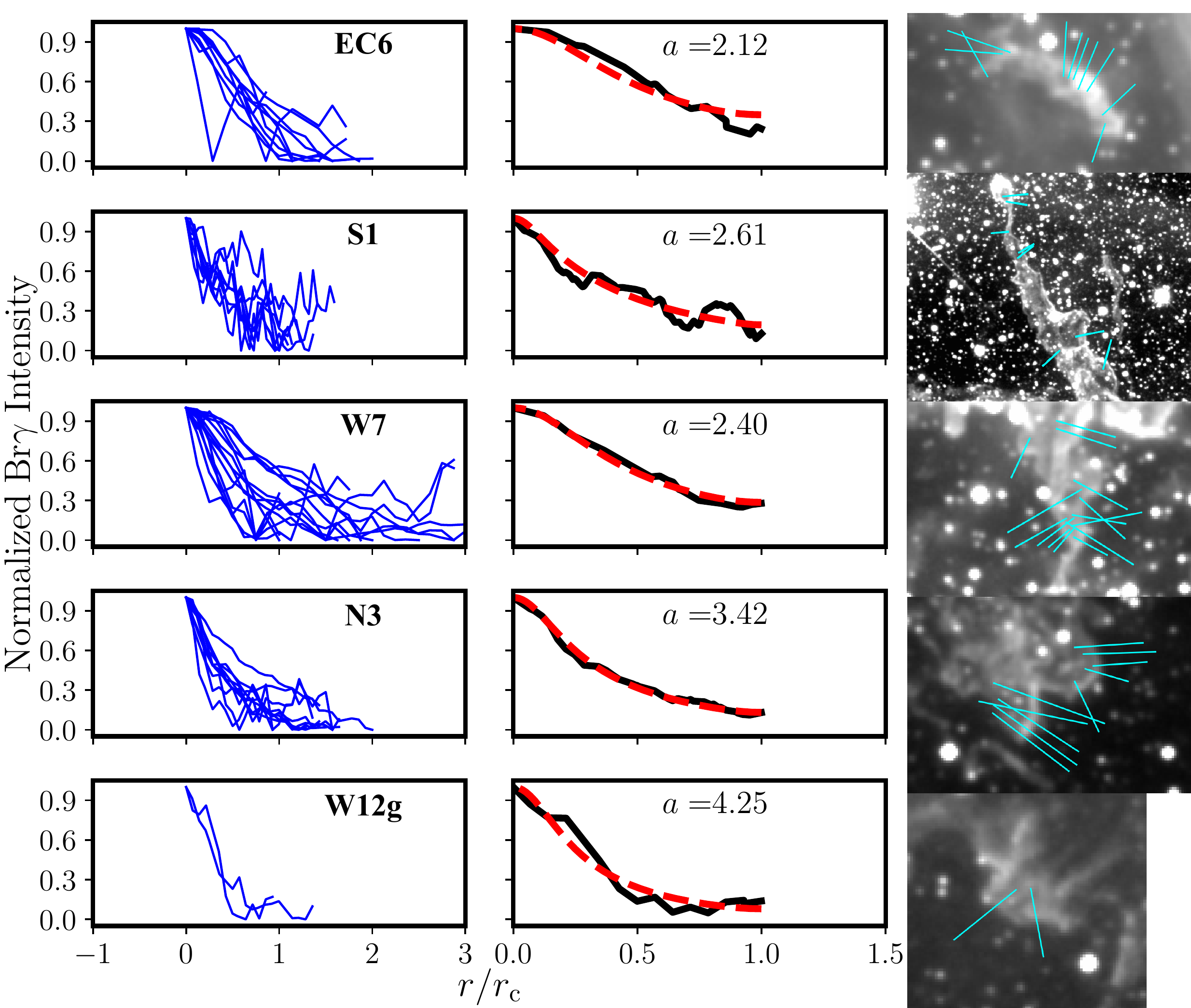}
\caption{{Continuum subtracted Br$\gamma$ profiles for five PDR interfaces in Carina.
The left column shows all Br$\gamma$ profiles for a given interface. The middle column shows the
median profile shape along with the best power-law fit. The right column displays an 
H$_2$ image of the object (not continuum-subtracted) together with the locations of all the
slices (shown in blue) that were used to fit the Br$\gamma$ profiles. The labels of the interfaces, as listed in the appendix, are indicated in the upper left.}}
\label{fig:prof_shapes}
\end{figure*}

\begin{figure}
\includegraphics[width=0.46\textwidth]{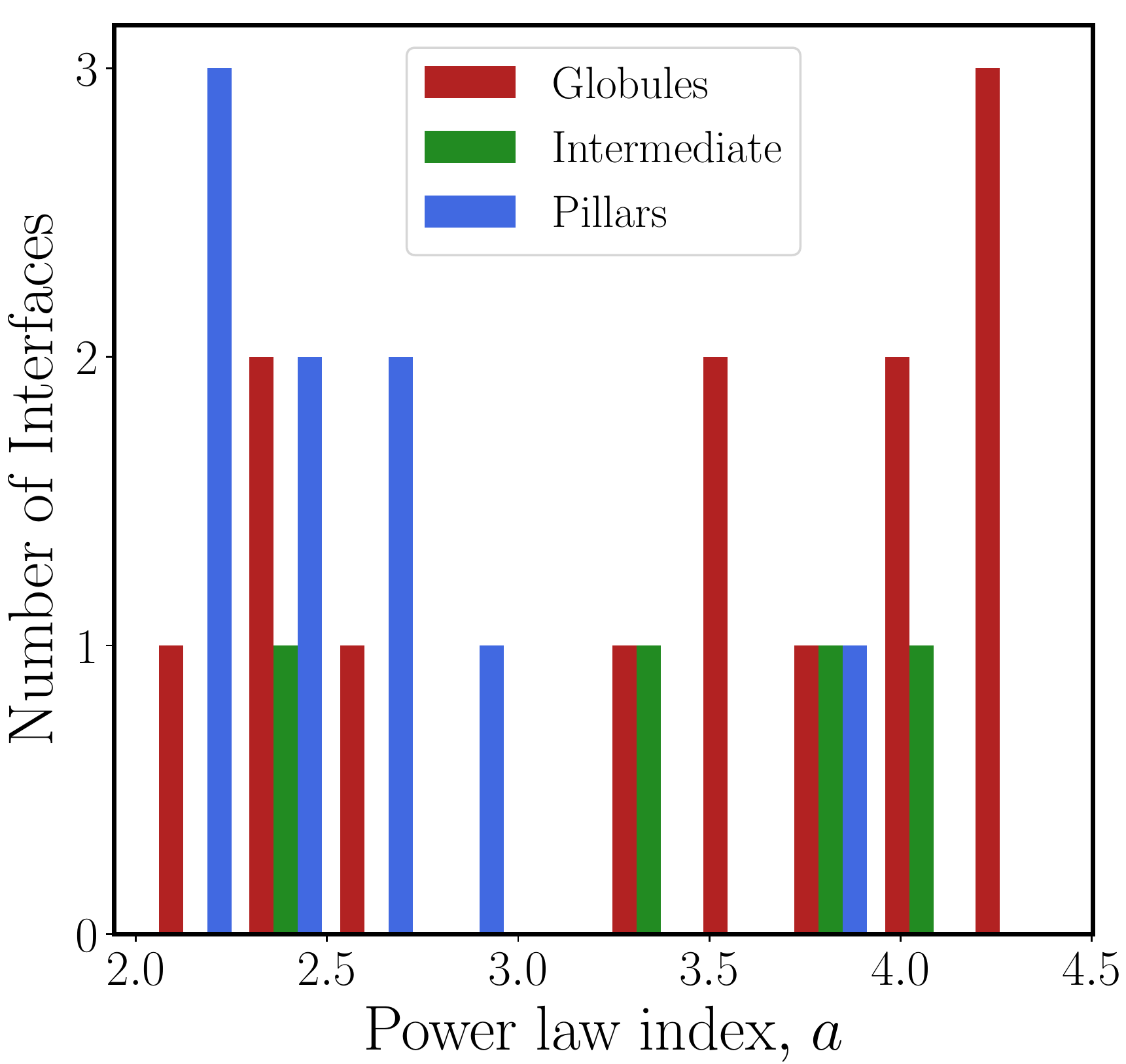}
\caption{Histogram of power law indices for 26 different interfaces in Carina and Cyg OB2, broken
into groups by their approximate geometric shape. }
\label{fig:powerlaw}
\end{figure}

\subsection{Br$\gamma$ Profiles and Interface Shapes}
\label{sec:brgProfiles}

We can study the IFs and photoevaporative flows in these regions
by measuring the spatial profiles of Br$\gamma$ intensity near the surface of the clouds.
Br$\gamma$ emission arises from photoevaporative flows off of the cloud.
Its brightness should reflect the density profile of the flow because Br$\gamma$ emissivity $\propto
n_e n_p$. Photoevaporative flows ablate from cloud surfaces at a few
times the sound speed \citep{bertoldi89}. If we assume that the flow moves with a roughly
constant velocity then conservation of mass requires that the density falls like $n\propto r^{-2}$ for
spherical clouds and $n\propto r^{-1}$ for cylindrical clouds, where $r$ is the distance from the
surface of the cloud. Assuming that the transition from neutral to ionized material is over a short distance, we then expect the Br$\gamma$ emissivity to follow a power law profile of the form:
\begin{equation}
\epsilon_{\rm Br\gamma} \propto \left(\frac{r+r_c}{r_c}\right)^{-b}
\label{eq:brg_emissivity}
\end{equation}

\noindent
where $r$ is the distance along the profile with $r=0$ corresponding to the peak in
Br$\gamma$ emission at the cloud's surface, $r_c$ is the radius of curvature of the cloud, and $b=2$ for a pillar and $4$ for a globule. Assuming that the photoevaporative flow extends outward from the cloud's surface to infinity, projecting the Br$\gamma$ emissivity (Equation \ref{eq:brg_emissivity}) onto the plane of the sky yields an observed Br$\gamma$ intensity profile of the form:
\begin{equation}
I_{\mathrm{Br\gamma}} = I_{\mathrm{max}}\left(\frac{r+r_c}{r_c}\right)^{-a}
\label{eq:brgProfile}
\end{equation}

\noindent
where $I_{\rm max}$ is
the peak Br$\gamma$ intensity on the profile and $a=b-1$. In other words, the effect of projection on the Br$\gamma$ profile is to make the power law shallower, decreasing the power law exponent by one, regardless of whether it is a pillar or globule. In the more realistic case that the photoevaporative flow does not extend to infinity and a uniform background takes over at some distance from the cloud, the projected intensity profile will not strictly be a power law since it goes to a constant at some radius. Assuming that the flow extends out at least a few times the radius of curvature of the cloud from the cloud's surface, the profile can be approximated as a power law with the exponent in between the unprojected and infinite flow cases. This means that we expect globules to have roughly power law Br$\gamma$ intensity profiles with exponent in between $-3$ and $-4$ and pillars to have exponents between $-1$ and $-2$\footnote{For instance, numerically projecting Equation \ref{eq:brg_emissivity} onto the plane of the sky for a flow that extends to $2r_c$ from the cloud's surface yields a projected Br$\gamma$ intensity profile of $I_{\rm Br\gamma} \propto r^{-1.78}$ for a pillar and $I_{\rm Br\gamma} \propto r^{-3.22}$ for a globule}.\par

To compare this with our data, we extract slices from the continuum-subtracted Br$\gamma$ images across cloud interfaces.
Slices were taken only on interfaces that appear edge-on, and we averaged the profiles over 5
pixels perpendicular to the slice to improve S/N. Slices must exclude stars, 
and this requirement limits the total number of slices.
Slices were taken on a variety of interfaces, including pillars, walls, and globules.
For a given interface, we extracted several independent slices and
aligned them by using the peak Br$\gamma$ emission. These slices generate a median profile
shape for each interface. Only the part of the slices on the HII region side of the Br$\gamma$ peak is kept for this analysis. To isolate the Br$\gamma$ emission in the photoevaporative flow from foreground/background nebulosity, a background level is subtracted from each slice. 
The background is quantified as the minimum pixel value along the slice. To make sure that the slices extend far enough out into the HII region that this minimum value does represent a background level and is not due to the photoevaporative flow itself, we further restrict slices to be longer than the radius of curvature of the object. 
 We grouped interfaces into three broad categories: globules, which have roughly spherical symmetry; pillars, which 
have approximately cylindrical symmetry; and intermediate shaped clouds, which are a mixture of the
two.  The pillars are specifically defined as those objects with a long axis being at least twice the length of a short axis.
 {For all the interfaces we can estimate a radius of curvature directly from the H$_2$ images by fitting an arc with
a constant radius of curvature to the feature.} The profiles were fit in the range $r=0$ to $r=r_c$.
To account for the blurring effect of seeing, the profile of Equation \ref{eq:brgProfile} is first convolved with a Gaussian of the width of the seeing before being fit to the data.\par

Figure \ref{fig:prof_shapes} shows these fits for five typical interfaces in Carina. The middle column shows the median
profile shape and the best fitting power-law. Images of the interfaces are also shown. The
top two interfaces are categorized as `pillars', the middle one as `intermediate', and the bottom
two as `globules'. The power law slope is steeper for the globules than the pillars,
as expected, because the extra curvature in the globules causes the gas to diverge over
shorter spatial scales than in the pillars. Doing this analysis for 26 objects in Carina and
Cygnus OB2\footnote{The interfaces in IC 1396 had large radii of curvature which made it impossible to get slices that were as long as the radius of curvature while still avoiding stars.} we derive the power law fits shown in Figure \ref{fig:powerlaw}. The
`globule' interfaces have an average power law index of $-3.45\pm0.22$ whereas the `pillar' interfaces have
an average index of $-2.54\pm0.16$. A two sample $t$-Test indicates that these means are significantly different with a $p$-value of 0.004. The globule power law exponents agree well with the expectation above that they fall in between $-3$ and $-4$, and the globule objects have steeper profiles than the pillars, also 
consistent with simple theoretical expectations. The power law exponents for the pillars are likely greater than the expected values between $-1$ and $-2$ because the objects are not exclusively cylindrical but consist of a mixture of cylindrical and spherical curvature.  A list of all the interfaces considered can be found in Appendix \ref{sec:appendix}

\begin{figure*}
\includegraphics[width=0.99\textwidth]{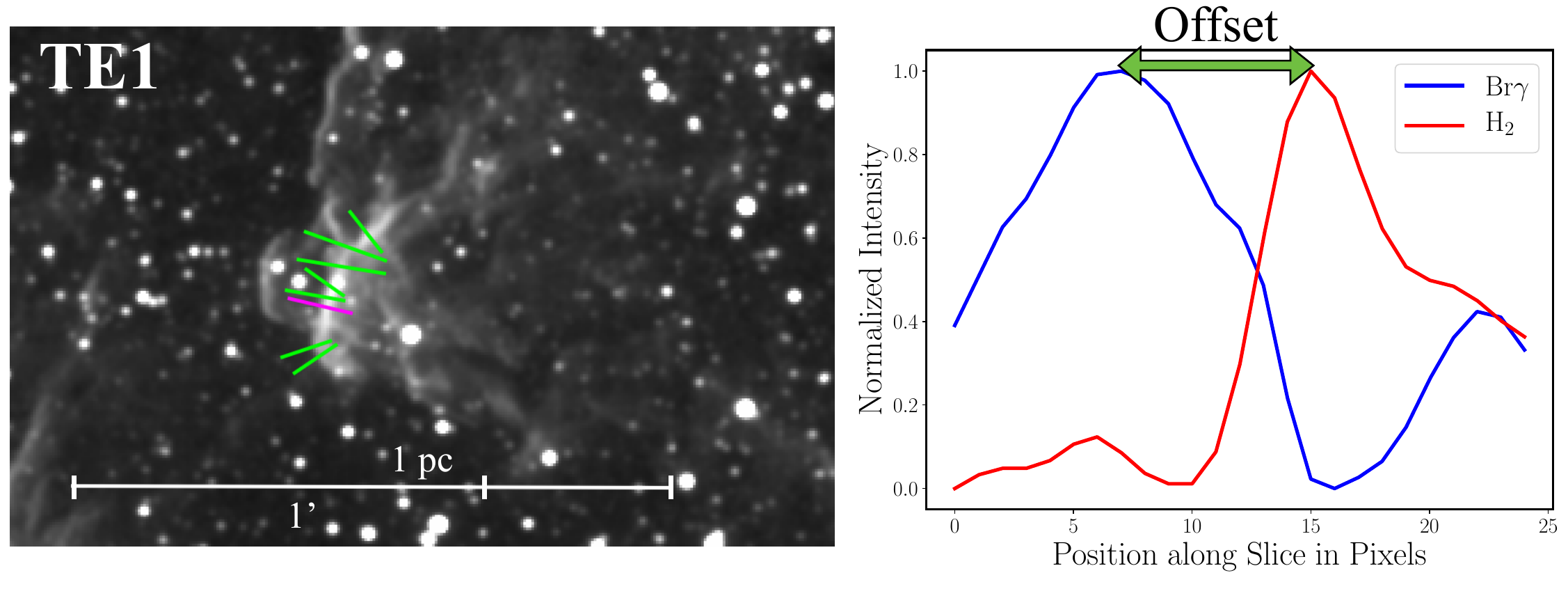}
\caption{{H$_2$ image (not corrected for continuum) of one of the objects in IC 1396
used for measuring spatial offsets between H$_2$ and Br$\gamma$ emission.
The green lines mark locations where the continuum-corrected 
spatial profiles were extracted for the emission lines. The spatial
offset between H$_2$ and  Br$\gamma$ is shown at right for the position
defined by the purple line. The interface's label, as given in the appendix, is indicated in the upper left.}
\label{fig:slice_ic}}
\label{example_slice_ic}
\end{figure*}

\begin{figure*}
\includegraphics[width=0.99\textwidth]{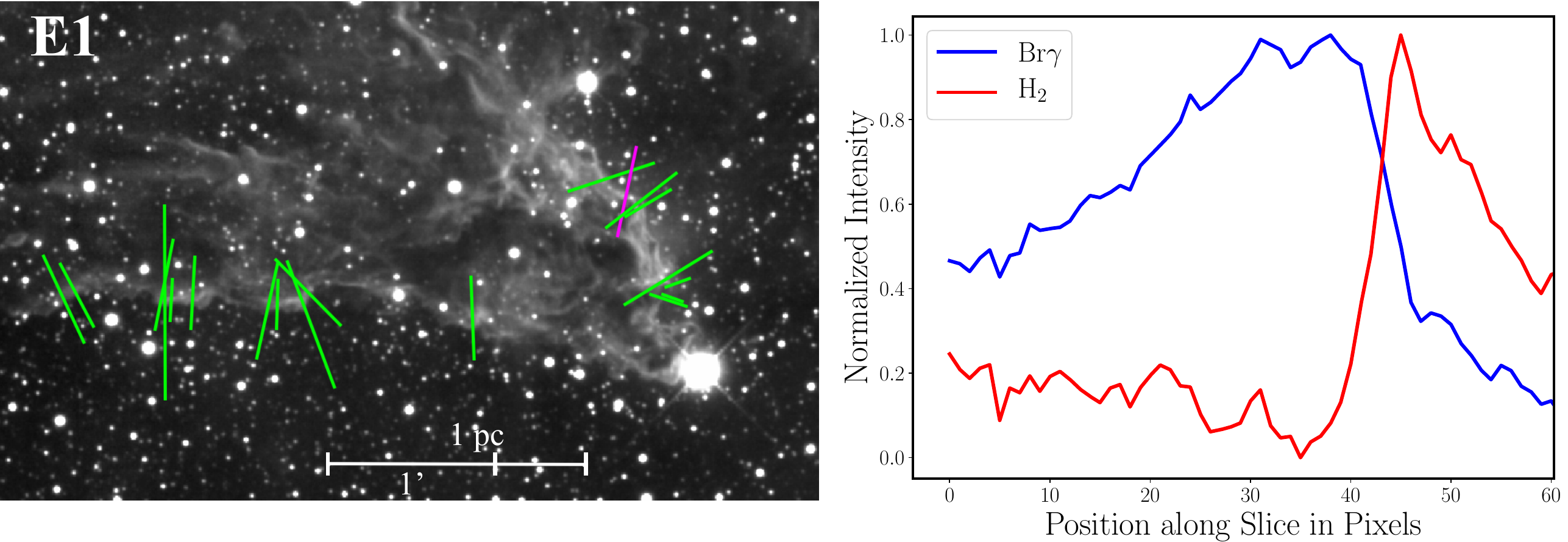}
\caption{{Same as Figure \ref{fig:slice_ic} for an object in Carina.}
\label{fig:slice_car}}
\end{figure*}

\begin{figure*}
\includegraphics[width=0.99\textwidth]{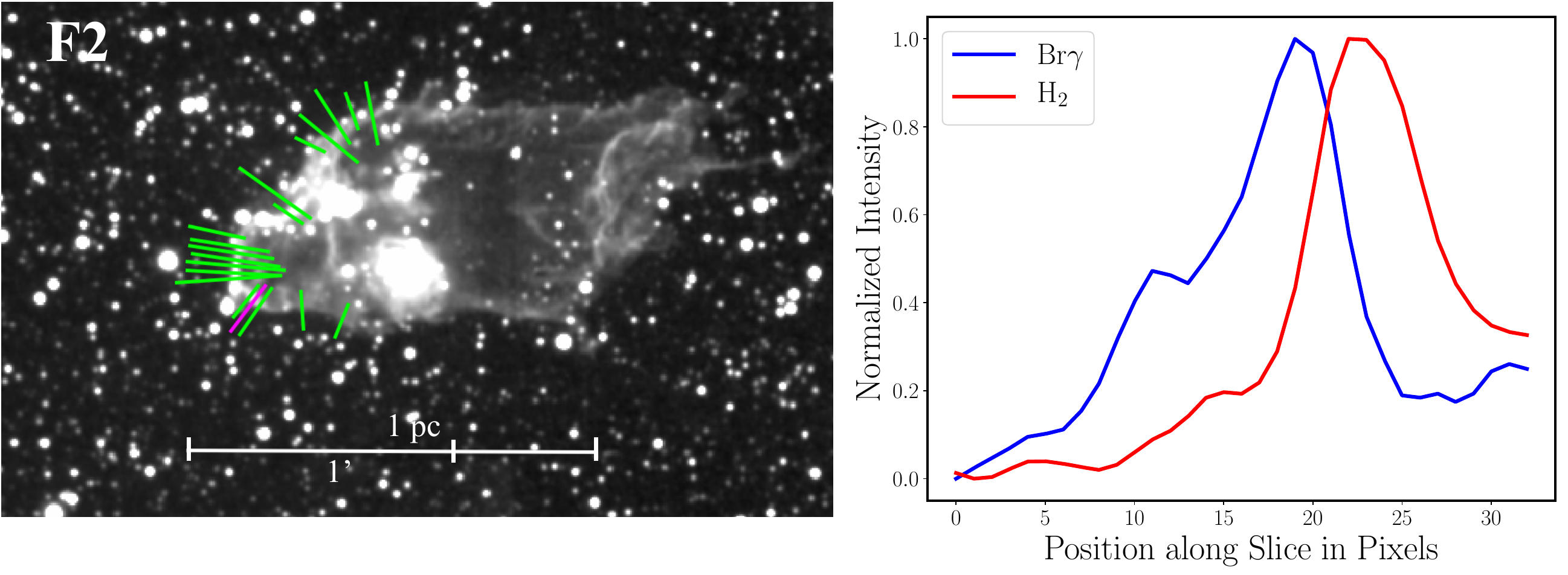}
\caption{{Same as Figure \ref{fig:slice_ic} for an object in Cygnus OB2.}
\label{fig:slice_cyg}}
\end{figure*}

\subsection{Spatial Offsets Between H$_2$ and Br$\gamma$}
\label{sec:offsets}
As noted in \citet{hartigan2015}, there is a distinct offset between the H$_{2}$ and Br$\gamma$ emission
when slices are extracted across PDR interfaces. Since the H$_{2}$ and Br$\gamma$ lines are so
close in wavelength, this offset should be independent of extinction, and it should reflect the physical
size of the PDR and ionization front. In this section we 
quantify this offset in Carina, IC~1396, and Cyg~OB2. Sample slices for each of the three regions are shown in Figs. \ref{fig:slice_ic}-\ref{fig:slice_cyg}, displaying typical offsets between H$_{2}$ and Br$\gamma$ for irradiated clouds.

Because the three star formation regions we are analyzing 
are at different distances, the images will have markedly 
different spatial resolutions even if the seeing were the same. 
To correct for this effect, we bin the IC 1396 and Cygnus OB2 data so that they have the same
pixel scale as the Carina data (Carina is the furthest region of the three).
We assume a distance of 1400~pc to Cyg OB2 \citep{rygl2012},  2300~pc to Carina \citep{smith2006}, and
870~pc to IC 1396 \citep{contreras2002}. At the distance of Carina, one pixel corresponds to $1.38\times10^{16}$cm. To equalize the
seeing, we blurred all the data down to the level of the image with the
worst seeing by convolving the images with Gaussians of varying widths. The image with the
worst seeing is the K-band Carina image with a FWHM seeing of $3.95\times10^{16}$cm (at the
distance of Carina), and all final images have this same level of smoothing. 

To quantify the observations we simply measure the spatial offset between
the peak emission in Br$\gamma$ and in H$_2$ in the smoothed images. 
Since the slices do not need to be longer than the interface's radius of curvature, we can consider more 
distinct objects than in Section \ref{sec:brgProfiles}. In total, we consider 33 objects across the three regions consisting of around 450 distinct slices. These objects are listed in Appendix \ref{sec:appendix}.\par

Figure \ref{fig:offset_3regions} shows the
cumulative distribution for physical offset sizes for the three regions.
IC 1396 appears to have the largest offsets while Cyg OB2 and Carina 
have roughly similar offsets. IC 1396 has median offset size of $9.9\times10^{16}$cm, while Carina has median offset size of $7.9\times10^{16}$cm, and Cyg OB2 has median offset of $7.8\times10^{16}$cm.
Figure \ref{fig:offset_3regions} shows that the offsets are all within roughly a factor of two of
$10^{17}$cm, despite the fact that the three regions span at least
three orders of magnitude in the intensity of ionizing radiation.

{We can use our data to determine if the observed offsets depend
upon the radius of curvature of the cloud or the strength of the radiation field.
Many interfaces have approximately spherical or pillar-shaped morphologies,
while others appear intermediate between these shapes. The H$_2$ image gives a characteristic radius of curvature for each interface, with a typical error
of $\pm$ 10\%. }To estimate the intensity of ionizing radiation incident onto each interface,
we use catalogs of the O and early B-type stars in the regions. For Carina,
we reference the catalog of O/B stars compiled by \citet{gagne2011}, and for Cygnus OB2, we apply
the list of \citet{wright2015}. Because HD~206267 is the lone O-star in IC 1396, we need
only consider that source.  We convert the spectral type to ionizing photon output 
using values from \citet{martins2005}. To approximate the flux of ionizing photons incident at
each slice location, we consider only the O/B stars that are visible to that slice
in the direction above the cloud interface. 
We combine the stars via:

\begin{equation}
F = \sum_i \frac{Q_i \cos(\theta)}{4\pi r^2},
\end{equation}

\begin{figure}
\includegraphics[width=0.45\textwidth]{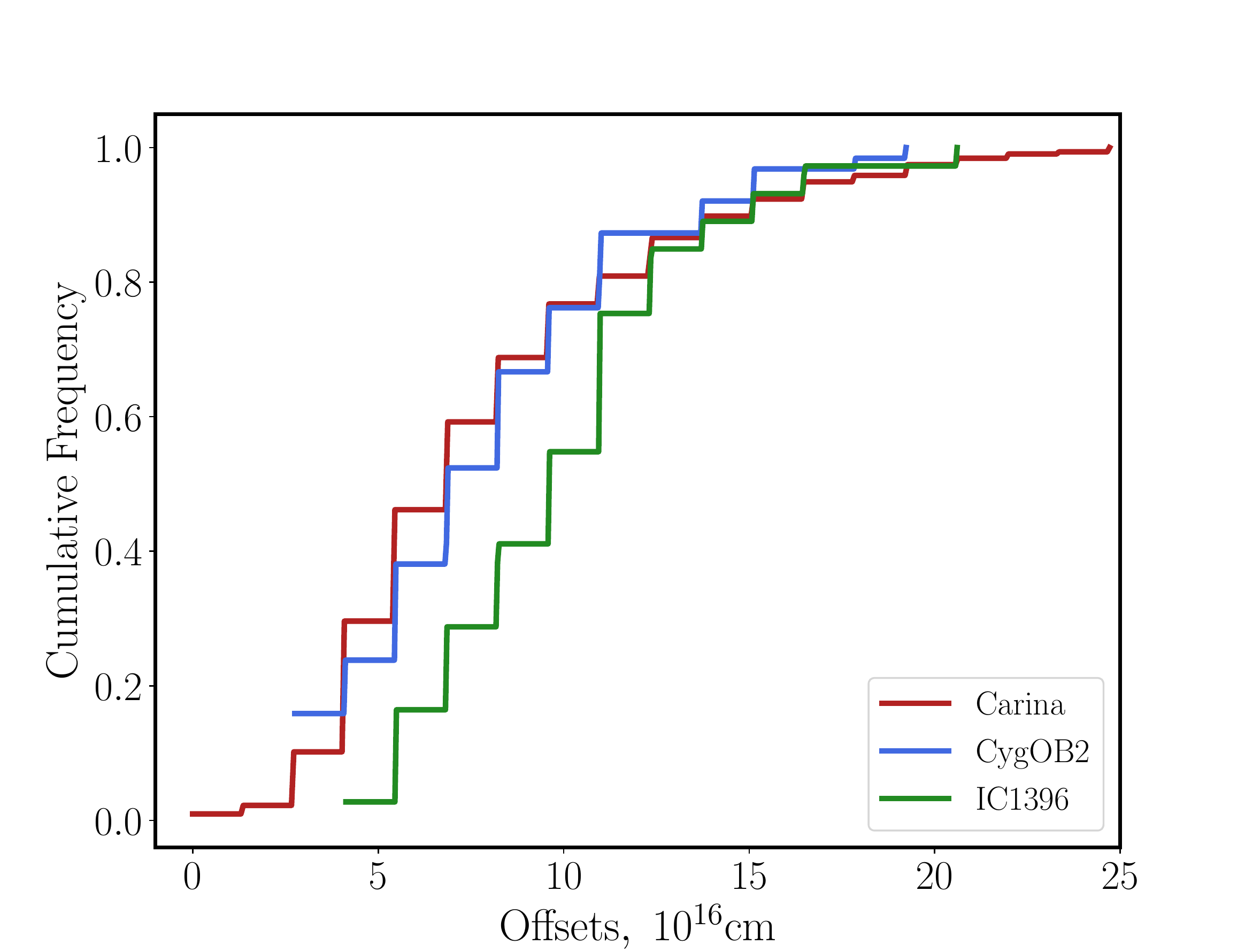}
\caption{Distribution of offset sizes. 
\label{fig:offset_3regions}}
\end{figure}

\begin{figure}
\includegraphics[width=0.45\textwidth]{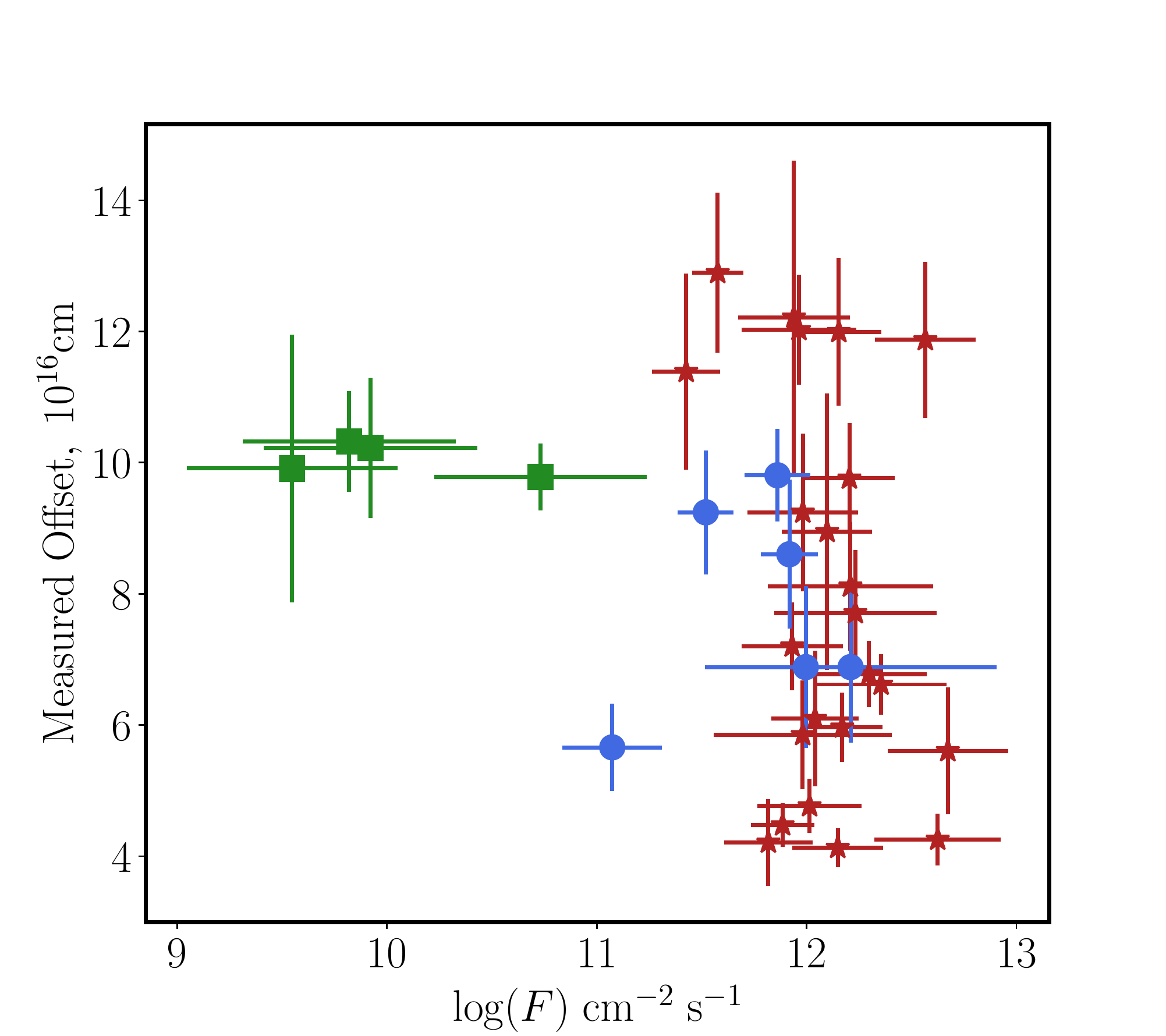}
\caption{Measured offset sizes versus the estimated incident flux of H-ionizing photons.
Objects in Carina are shown in red, those from Cygnus OB2 are in blue, and those from IC 1396
are in green. The flux of ionizing photons spans three orders of magnitude.}
\label{fig:o_v_f}
\end{figure}

\begin{figure}
\includegraphics[width=0.45\textwidth]{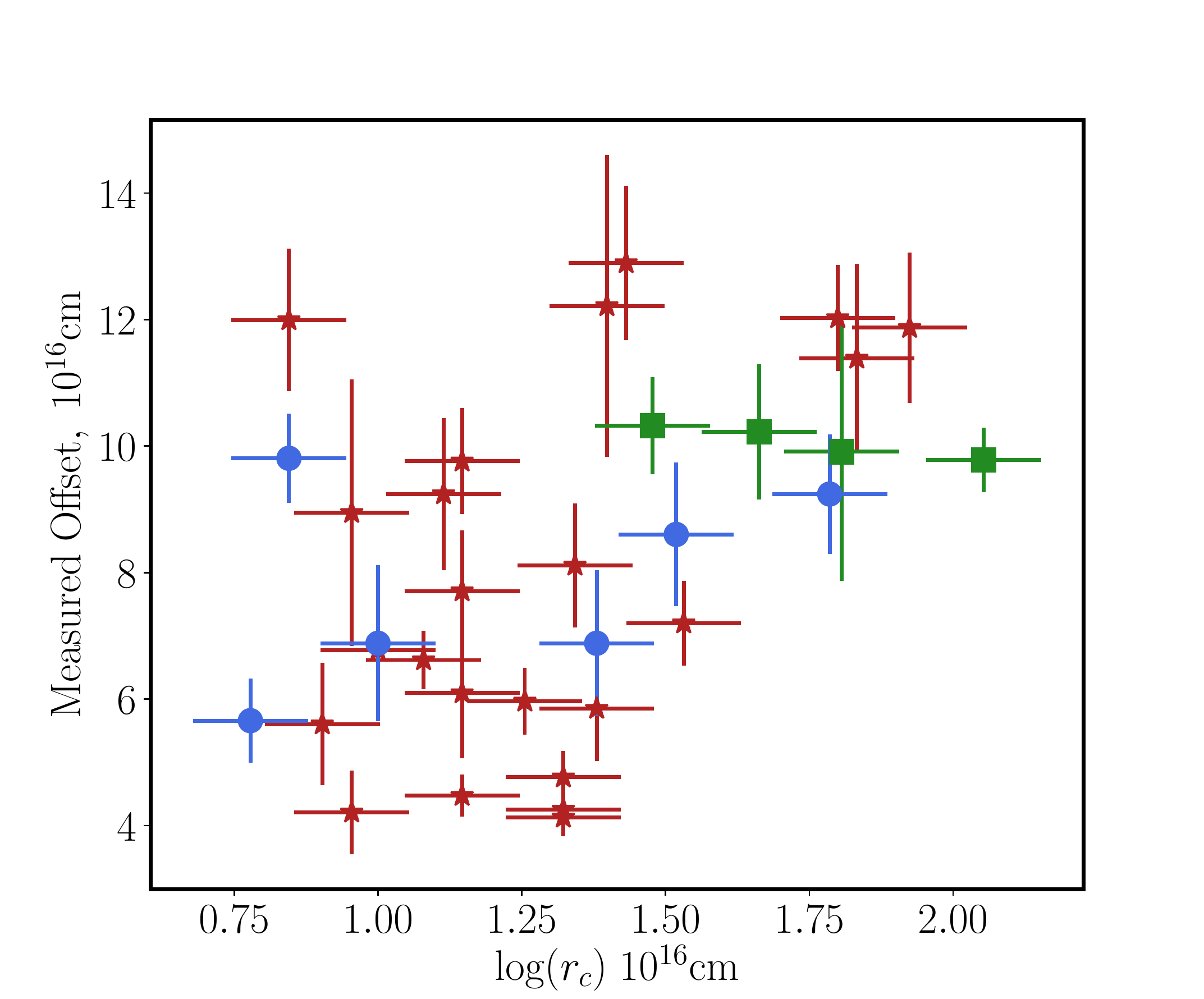}
\caption{Measured offset sizes versus the measured radius of curvature. Objects in Carina are shown in
red, those from Cygnus OB2 are in blue, and those from IC 1396 are in green.}
\label{fig:o_v_rc}
\end{figure}

\begin{figure}
\includegraphics[width=0.46\textwidth]{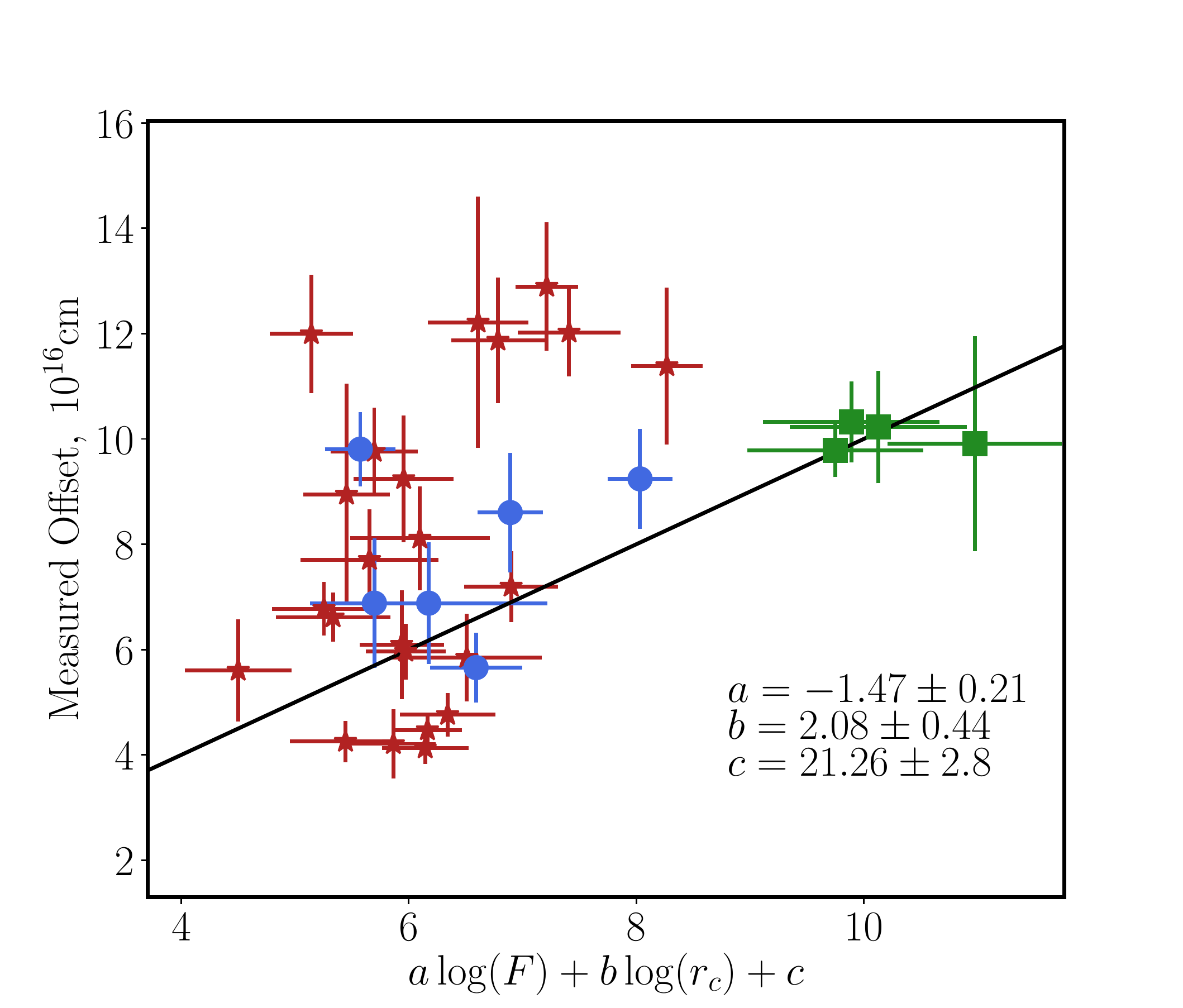}
\caption{Edge-on view of best fitting plane to the measured offset values along with best-fitting parameter
values. Objects in Carina are shown in red, those from Cygnus OB2 are in blue,
and those from IC 1396 are in green.}
\label{fig:plane_fit}
\end{figure}

\noindent
where $Q_i$ is the ionizing photon output of the $i$-th star, $\theta$ is the angle
between the line connecting the star to the interface and the normal to the interface
(which is taken to be the slice we make), and $r$ is the projected distance
between the star and the interface.

To account for projection effects, we calculate an uncertainty in
the flux, $F$, by assuming that the distance, $r$, is known
to within 25\%\footnote{Which corresponds to the vector between the star and interface being within 37$^{\circ}$ to the plane of the sky.}. This uncertainty is then propagated through to the final flux value
for each slice. We average the flux values for each slice on an object to get an estimate for the flux value for that object. Since the uncertainties for the flux value at each slice on an object are not uncorrelated, we just take the rms of these uncertainties as an estimate for the uncertainty in the ionizing flux for that object. { It is also possible to estimate the FUV flux at each point in the regions from the observed far-infrared (FIR) intensities \citep[e.g.][]{kramer2008, rocc2013} but since the EUV flux is more relevant to studying the IFs and photoevaporative flows, we do not explore that method here.}

The offsets measured from slices
on an object are averaged to give a mean offset for that object and a
standard error of the mean.

After these measurements are completed, each object has an associated radius of
curvature, an ionizing flux level, and a spatial offset between Br$\gamma$ and H$_2$.
Figure \ref{fig:o_v_f} shows the measured offsets as a function of the estimated
incident flux. The range in $\log(F)$ shows that our objects differ in the level of
incident flux by around three orders of magnitude between the weakest irradiation environment in IC
1396 and the strongest in Carina. Figure \ref{fig:o_v_rc} shows the offsets as a function of
the measured radius of curvature. There is a reasonably strong positive correlation between
the offset size and the radius of curvature and a very weak negative correlation with respect to
the incident flux. A Spearman rank-order test for Figure \ref{fig:o_v_f} gives a $p$-value of no correlation of 0.05 and a $p$-value of 0.007 for Figure \ref{fig:o_v_rc}. The low $p$-value in Figure \ref{fig:o_v_f} is probably driven by the IC 1396 interfaces at low incident flux. Not including the IC 1396 interfaces yields $p$-values of 0.3 and 0.07 for Figures \ref{fig:o_v_f} and \ref{fig:o_v_rc}, respectively.
 Both Figures \ref{fig:o_v_f} and \ref{fig:o_v_rc} show a significant spread in the
offset sizes even after all the individual slices for an object have been
averaged. Each point in Figures \ref{fig:o_v_f} and \ref{fig:o_v_rc} is already an
average of 10-20 individual slices.

The spatial offsets between Br$\gamma$ and H$_2$ should depend jointly upon both the incident
flux and radius of curvature, which introduces scatter to Figures \ref{fig:o_v_f} and
\ref{fig:o_v_rc} when offsets are plotted against either $\log(F)$ or
$\log(r_c$) individually. To correct for this effect, we fit a plane to the offset values of the form:

\begin{equation}
\mathrm{O} = a\log(F) + b\log(r_c) + c
\end{equation}

\noindent
where $\mathrm{O}$ is the measured offset. Figure \ref{fig:plane_fit} shows an edge-on view of this plane
along with the best fit parameters $a,\; b,$ and $c$.  The scatter in the planar view 
(Figure~\ref{fig:plane_fit}) is less than that for $\log(F)$ (Figure~\ref{fig:o_v_f}) and
for $\log(r_c)$ (Figure~\ref{fig:o_v_rc}), though considerable intrinsic scatter remains,
caused in large part by the complex geometrical shapes of the individual objects and uncertainties in the true (deprojected)
distances of the ionizing sources from the cloud interfaces. 
Nevertheless, there is a weak positive trend, $b = 2.08\pm 0.44$, of the offsets with radius of curvature
of the cloud and a weak negative trend for the offset with the incident flux,
$a = -1.47\pm0.21$.

\section{Photoionization Models}\label{sec:models}

\subsection{Motivation}

{Our two primary motivations for developing a numerical model of these regions are (1) to explain the
differences and similarities in the offset sizes observed in the three regions and (2) to explore the relevant
physical processes that define the size of the offsets.} The fact that the three regions show
similar offsets indicates that offsets might end up being more useful as a rough distance
indicator than as a PDR diagnostic tool.
Hence, rather than use the spatial offset between Br$\gamma$ and H$_2$ 
to determine physical conditions, the focus will be to understand
the underlying processes that create these offsets and to predict any correlations with other observables.

\begin{figure*}
\centering
\includegraphics[width=.92\textwidth]{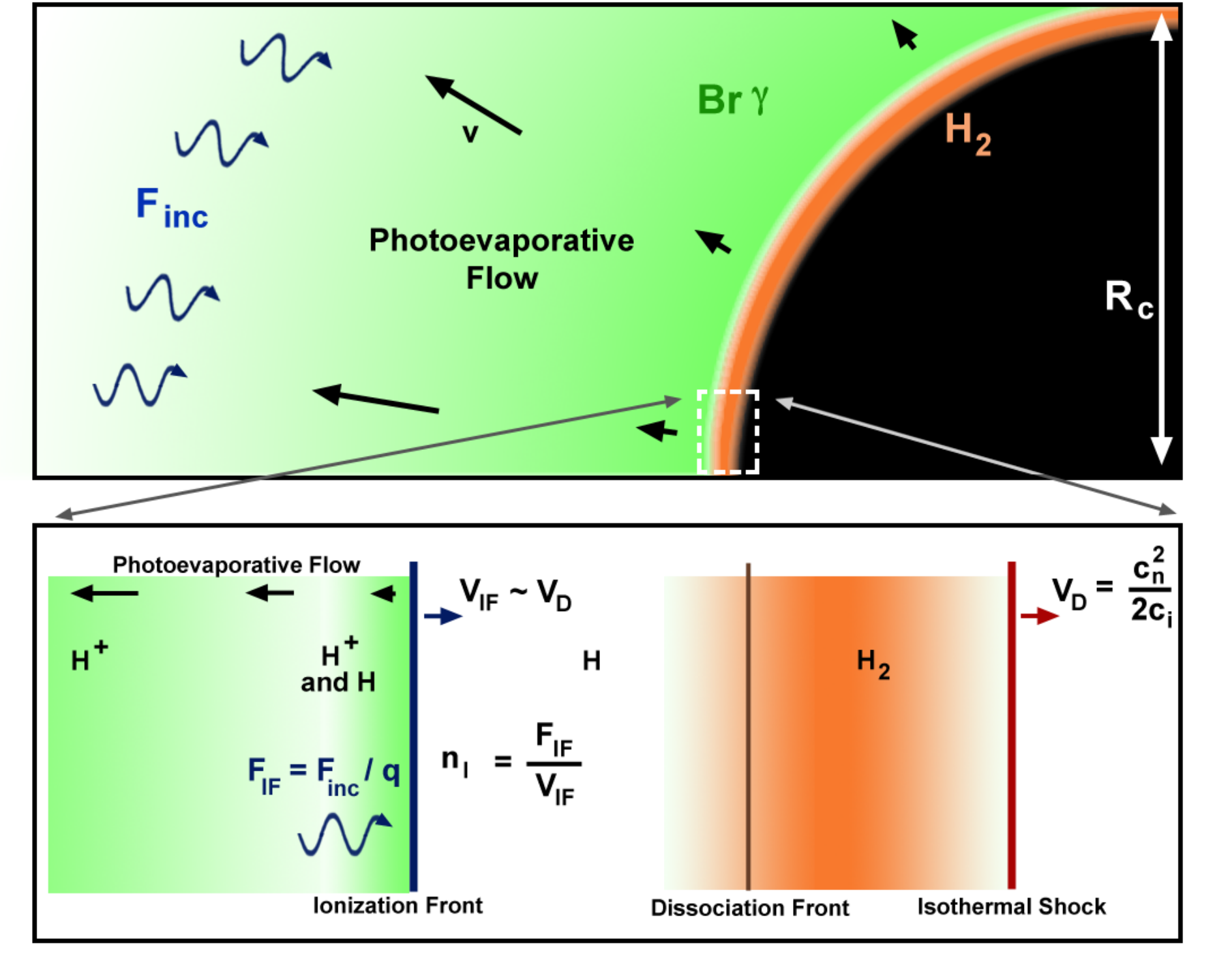}
\caption{Schematic of a photoevaporative flow from an irradiated globule.
Top: Ionizing flux $F_{\rm inc}$ from the left impinges upon a dense spherical globule
of radius $r_c$ and drives an outflow. The green colors depict
Br$\gamma$ emission and the red show H$_2$ emission. Bottom: Recombining gas in the
outflow reduces the incident ionization flux by a factor of $q$ at the ionization front.
The ionization front progresses into the cloud at a velocity $v_{\rm IF}$ $\sim$ $v_{D}$,
where $v_{D}$ is the velocity of the isothermal D-shock, set by the
sound speeds $c_i$ and $c_n$ in the ionized and neutral gases, respectively.
The model solves ionization and dynamical equations
for $q$ to determine $n_{\rm I}$, which in turn becomes an input parameter to
\textsc{Cloudy} for determination of the H$_2$ emission. The velocity arrows
depicted in the figure are relative to the unperturbed globule.}
\label{fig:model_setup}
\end{figure*}

\begin{figure*}
\includegraphics[width=0.95\textwidth]{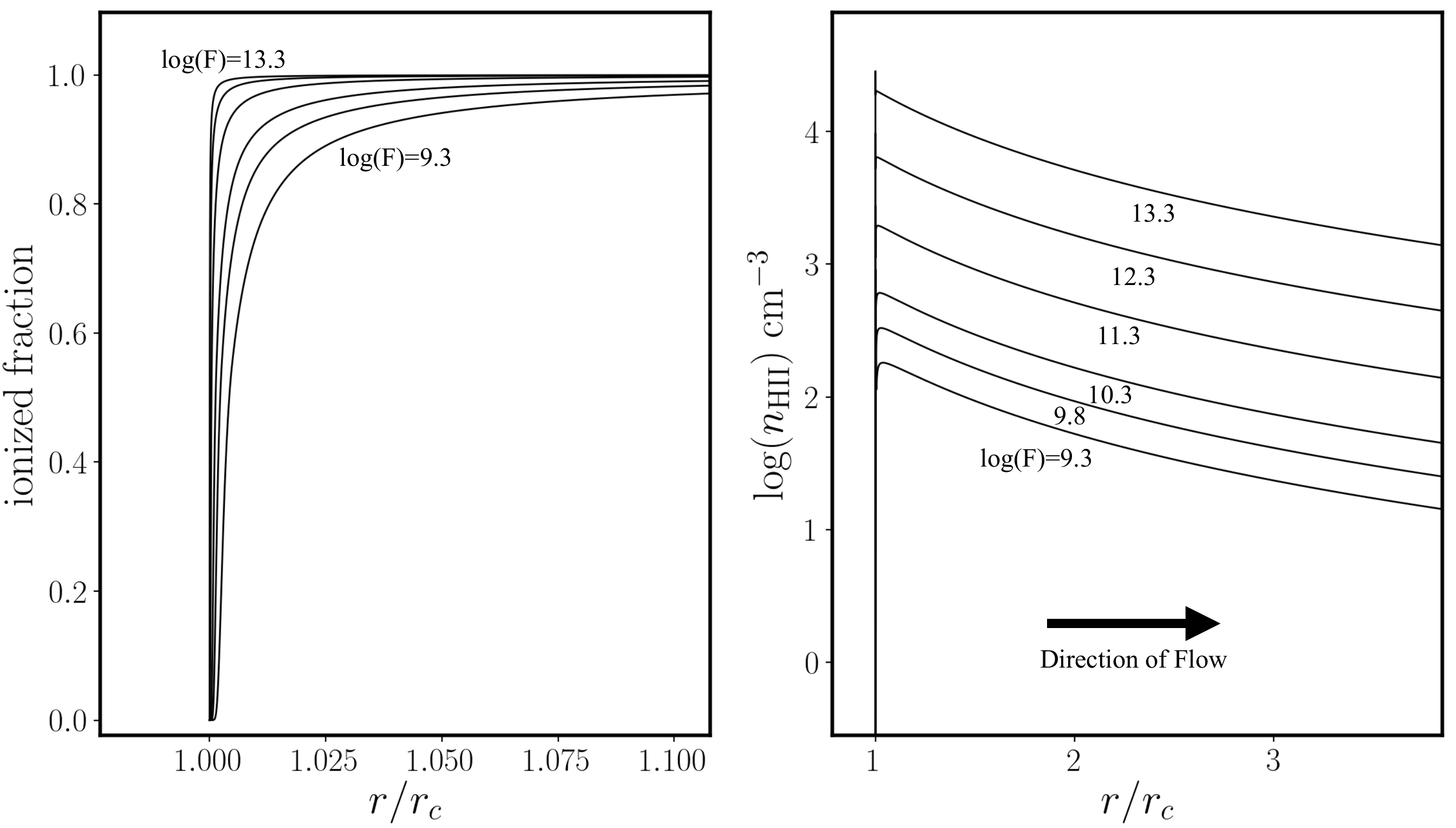}
\caption{Model results for models with $r_c = 5\times10^{17}$cm. The left plot shows profiles of
the ionization fraction. The right plot shows the electron number
density for the same models. In order of increasing ionization fraction at a given radius,
the models have $\log(F) = 9.3, 9.8, 10.3, 11.3, 12.3, 13.3$ cm$^{-2}$s$^{-1}$. }
\label{fig:model_1}
\end{figure*}

\subsection{Model of a Photoevaporative Flow} 

The Br$\gamma$ emission originating near the molecular clouds comes from a photoevaporative flow
of ionized gas off the cloud surface. Therefore, to accurately model the offsets and their
dependence on various parameters, a rough model of a photoevaporative flow is necessary.
We follow \citet{bertoldi89} and solve the equations of hydrodynamics and ionization balance in a simplified
spherically symmetric, 1-D geometry and in the steady-state. In this formalism, all partial time derivatives
in the fluid equations are set to zero. A picture of the physical situation along with a definition of the variables is shown in Figure \ref{fig:model_setup}. We solve the continuity equation, momentum equation,
and equations for ionization balance as:

\beq
\frac{d}{dr}\left(r^2n_{\rm H}v\right) &= 0 \\
v\frac{dv}{dr} &= -\frac{1}{\rho}\frac{dP}{dr} \\
\frac{dF}{dr} &= (1-x) n_{\rm H}\sigma_\nu F \\
\frac{dx}{dr} &= \frac{1}{v}\left(-\alpha_b x^2 n_{\rm H} + (1-x)\sigma_{\nu} F\right)\\
\eeq

\noindent
where $n_{\rm H}$ is the hydrogen number density, $\rho = m_{\rm H} n_{\rm H}$, $P$
is the pressure, $v$ is the velocity of the flow, $F$ is the flux of
ionizing photons at a certain point in the flow, $x$ is the ionization fraction, $\alpha_b$
is the case-B recombination coefficient, and $\sigma_{\nu}$ is the frequency averaged
photoionization cross section for hydrogen. We use $\alpha_b=2.7\times10^{13}$cm$^{3}$s$^{-1}$
and $\sigma_{\nu} = 3.0\times10^{-18}$cm$^{2}$. Following \citet{lefloch1994},
instead of solving a separate energy differential equation, we write the sound speed 
as:

\beq
c^{-2} = c_i^{-2} + \left(c_n^{-2} - c_i^{-2}\right)\left(1-\sqrt{x}\right)^2
\eeq

\noindent
which is essentially an arbitrary interpolation function between the two extremes, $c_n$ and $c_i$, 
the isothermal sound speeds for the cold neutral gas and the warm ionized gas,
respectively. We take $c_n$ = 0.8 km/s corresponding to $T=100$~K neutral hydrogen and $c_i$=11.4 km/s
corresponding to $T=10^{4}$~K ionized hydrogen. As discussed in \citet{lefloch1994},
writing the sound speed explicitly as such requires that the flow be in thermal equilibrium. More specifically, it requires that
the thermal timescale be much less than the dynamical timescale of the flow. \citet{lefloch1994} argue
that this is the case for flows with strong irradiance where the flow acts as
an insulating boundary layer and absorbs most of the incoming ionizing photons before they
reach the IF. This will be the case for the regions we consider with strong irradiance in Carina and Cygnus but
might not be accurate for the objects in IC 1396 which have fairly weak incident radiation. 

We assume the ionization front is D-type and is therefore preceded by a shock wave
in the steady-state (Figure \ref{fig:model_setup}). Following \citet{bertoldi89} we model the flow from the neutral part outwards.
\citet{bertoldi89} found that the velocity of the IF
is approximately the D-critical velocity under a
wide range of conditions. We write the IF velocity as  

\beq
v_{\rm IF} = \phi_d v_D
\eeq

\noindent
where $v_D$ is the D-critical velocity $\approx c_{n}^2/2c_i \approx 0.03$ km/s \citep{spitzer1968} and $\phi_d$ is an order unity number. 
We will assume that $\phi_d = 1$. Additionally,
the density of the shocked gas layer will be 

\beq
n_{\rm I} = \frac{F_{\rm IF}}{v_{\rm IF}}
\eeq

\noindent
where $F_{\rm IF}$ is the flux of ionizing photons that actually reaches the IF. We
parameterize the effect of shielding by the flow by writing $F_{\rm IF} = F_{\rm inc}/q$
where $q$ is unknown \textit{a priori} and $F_{\rm inc}$ is the incident flux. The model
starts in the neutral part of the cloud with $F_0 = 10^{-8}F_{\rm inc}$ and $x_0 = 10^{-8}$
and numerically integrates the equations as it moves outwards. The gas accelerates as it
heats up and becomes super-sonic. However, since we are interested in the density profiles near
the IF and not further out along the flow, we do not follow the super-sonic
evolution. Once the gas becomes faster than $0.99$ times the sound speed, we set the gas velocity as the
sound speed and do not consider the momentum equation any further. This allows us
to avoid the complication that the momentum equation is actually singular at the sonic point.
The models are integrated from $1r_c$ to $5r_c$. The parameter $q$ is chosen so that
the flux at $5r_c$ is within 5\% of the intended incident flux, $F_{\rm inc}$. We
repeat the process for various radii of curvatures and fluxes relevant to our regions. Ionization fraction
and density profiles for models with $r_c = 5\times10^{17}$cm for a variety of incident ionizing fluxes are shown in Figure \ref{fig:model_1}. \par

The models show a spike in the density of HII in the profiles in the ionization fronts.
This occurs because there is a large density contrast across the IF of roughly a
factor of $c_i/v_{\rm IF} \approx 200$. This causes the IF to have higher density than
the photoevaporative flow and so even if the IF is not completely ionized, it can
have a higher HII density than in the flow. {This spike is model dependent (it is sensitive to our simplified energy equation) and it is unclear if it would be present if heating and cooling are explicitly considered. We do not consider this here as the spike is too narrow to affect the observed offset between Br$\gamma$ and H$_2$. It will be completely smoothed once blurred by the seeing.} \par 

\begin{figure}
\includegraphics[width=0.48\textwidth]{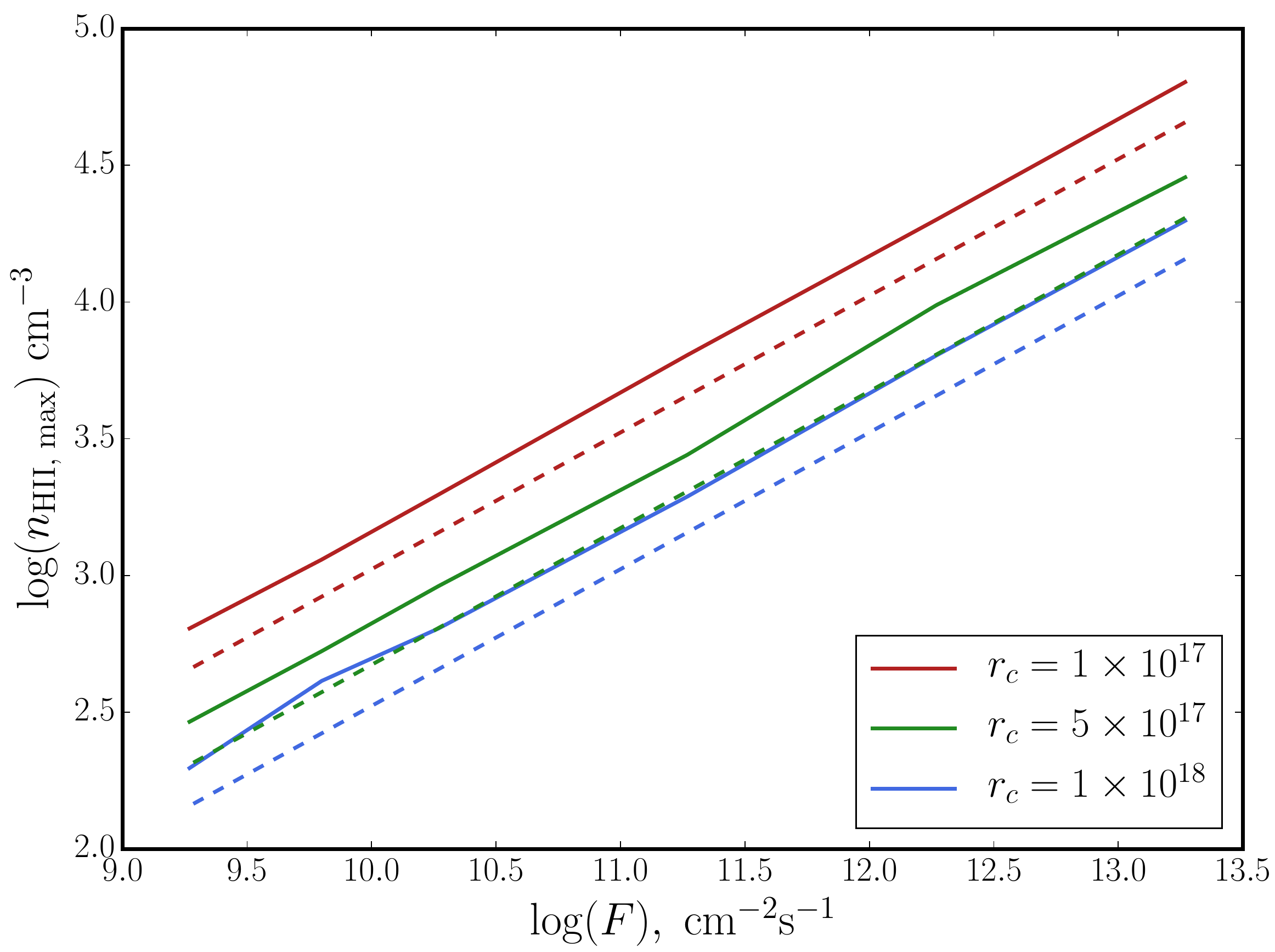}
\caption{Solid lines show the maximum HII density in the simulations as a function of $F$
and $r_c$. In dotted lines is shown the expectation for a photoevaporative flow with constant
velocity and infinitely thin IF (cf. Equation \ref{eq:ana}).  }
\label{fig:model_2}
\end{figure}

Figure \ref{fig:model_2} shows the scalings of
the maximum HII density in the flow with the flux and radius of curvature. Also
shown is the expectation for a flow with constant velocity and infinitely thin IF given
by \citep[e.g.][]{tielens05}:

\beq
n_{\mathrm {HII,\; max}} = \left(\frac{3 F}{r_c \alpha_b}\right)^{1/2}
\\
\label{eq:ana}
\eeq

This density is found by assuming that the photoevaporative flow acts as an insulating boundary
layer that absorbs most of the ionizing photons before they reach the IF. We
can see that the scalings in the simulations with regard to $F$ and $r_c$ are
the same as Equation \ref{eq:ana} but that the simulation densities are roughly $1.5\times$ higher. This
is due to the fact that the ionization fronts have non-zero thickness. Equation \ref{eq:ana} assumes
that all of the ionized hydrogen lies outside the IF and moves at
a constant velocity where, {in our models}, the peak density of HII occurs inside the IF where
gas is not completely ionized but is very dense because the gas is still moving
slowly. As mentioned above, conservation of mass requires that the density of the gas before
the IF is roughly $\approx 200$ times that of the gas on the ionized side
of the IF. Within the IF, the gas density will be somewhere in between. \par

This scaling of $n_{\mathrm {HII,\; max}}\propto F^{1/2}$ is found to agree quite well with the pillar mass loss rates in Carina, M16, and NGC 3603 analyzed in \citet{mcleod2016}.

\begin{figure*}
\includegraphics[width=0.95\textwidth]{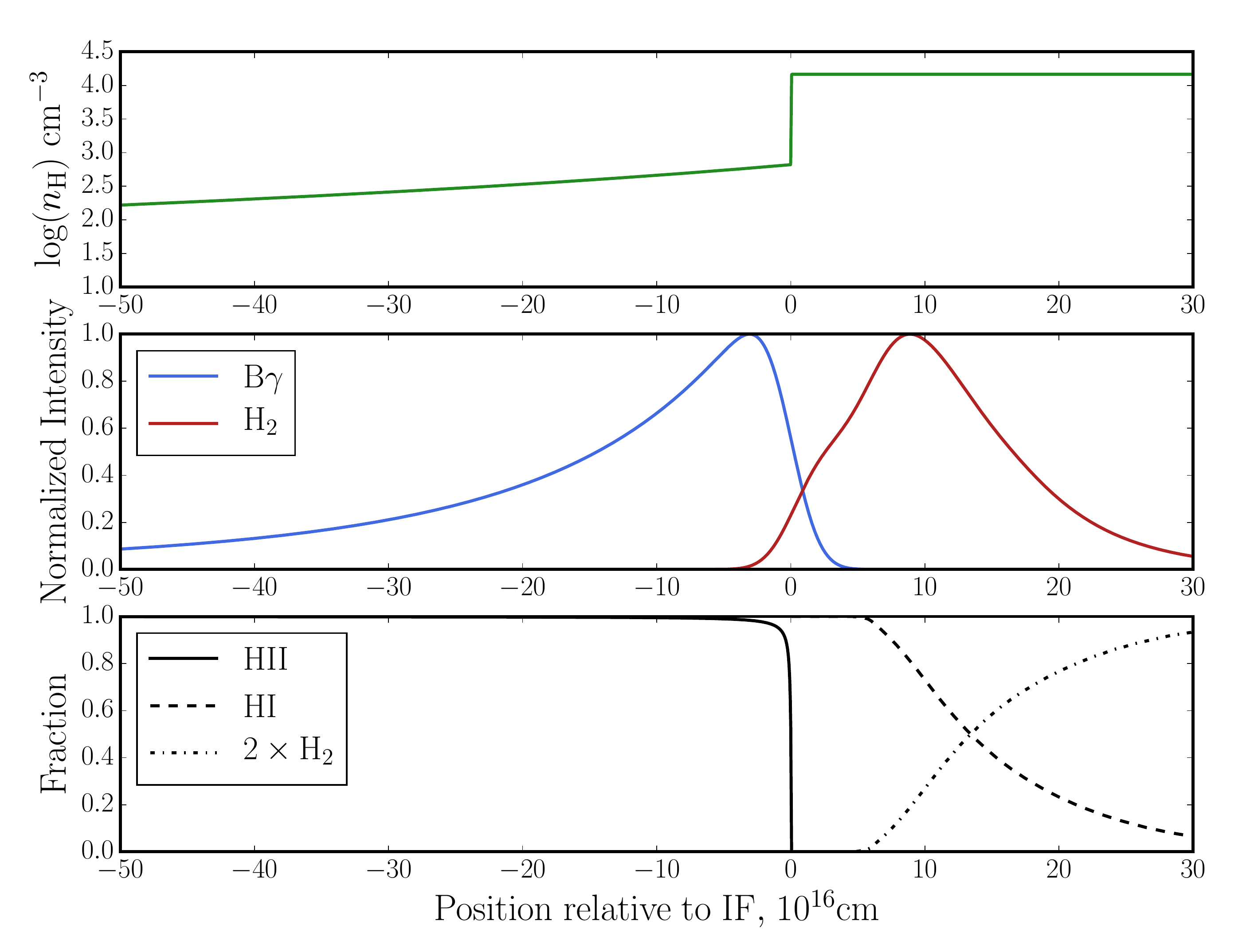}
\caption{Sample model result with $r_c =5\times10^{17}$cm and $F=10^{10.2}$cm$^{-2}$s$^{-1}$.
The profiles are centered on the ionization front defined as where the ionization fraction is
50\%.  Note that we do not see a rise in the neutral HI fraction right
after the IF because this is where the dynamical model stops and the \textsc{Cloudy} model
begins. The \textsc{Cloudy} model starts with the HI fraction being $1$.  }
\label{fig:model_sample}
\end{figure*}

\subsection{Synthetic Br$\gamma$ and H$_2$ Spatial Line Profiles}
{Synthetic observations play a key role in interpreting real observations. For a recent review of applications of synthetic observations, see \citet{haworth}. In this investigation, we generate synthetic spatial emission profiles from the
photoevaporative flow models described above.} We create the profile for each line separately
and then stitch them together. To create the Br$\gamma$ profile, we simply square the HII
density profile calculated in the model. Because we are not interested in the actual Br$\gamma$ emissivity, we
do not need to worry about the recombination coefficient. In doing this, we ignore
the dependence of the recombination coefficient on the temperature but that is a minor
effect compared to the variations in HII density.

To create the H$_2$ profile, we use
the photoionization code \textsc{Cloudy}, version 13.03, described by \citet{ferland2013}. \textsc{Cloudy} works
by providing it a density law and input spectrum and it does the necessary radiative
transfer to predict the H$_{2}$ emissivity. \citet{bertoldi1996} found that, after a transitory initial phase,
the dissociation front (DF) is trapped within the shocked layer. This means that the PDR,
where the HI exists and where the 2.12$\mu$m emission is coming from, is situated in the
shocked layer. Hence, we model the PDR in \textsc{Cloudy} by assuming a constant density of
$n_{\rm I}$ found from the dynamical models. The shape of the input spectrum is that
of an O6 star from the TLUSTY stellar atmosphere models \citep{hubeny1995} with all of the
Lyman continuum photons removed. This is only an approximation as the photoevaporative flow will
modify the spectrum more than just extinguishing the Lyc photons, however it should be enough
to understand general trends in the offsets.

The models are also approximate in the sense
that \textsc{Cloudy} does static modeling but the PDR is actually evolving as hydrogen advects
across it at the speed of the dissociation front. \citet{bertoldi1996} find that non-equilibrium effects in
the PDR, including transient photochemistry, are important when the incident flux and the density of
the cloud is low. From their Figure~4, the regions we consider that have strong incident
flux like that characteristic of Carina and Cygnus should be well-matched with equilibrium PDR models (like \textsc{Cloudy}) though objects with
weaker incident flux like in IC 1396 might not be. Despite this limitation, we use \textsc{Cloudy} to get a rough
understanding of the general trends in the offsets. 
The models run deep into the molecular region of the clouds, and we adopt \textsc{Cloudy}'s default
chemical abundances for an HII region, which are average values of the
abundances in the Orion Nebula \citet{rubin1991,baldwin1991,osterbrock1992,rubin1993}, with
a metallicity of 0.4 solar. The models also include dust grains with
the size distribution and abundances of grains found in the Orion Nebula. Additionally, the models
include a cosmic ray flux characteristic of the galactic background taken from \citet{indrilio2007}. We combine
the two profiles together simply by stitching the H$_2$ profile onto the location
where the dynamical models begin (with $F_0 = 10^{-8}F_{\rm inc}$ and $x_0 = 10^{-8}$) . Both the Br$\gamma$ and
H$_2$ profiles are then blurred to the same level of seeing as the NIR
images by convolving them with the appropriate Gaussian. A sample model result is shown in
Figure \ref{fig:model_sample}. The density profile shows the $1/r^{2}$ fall-of in the photoionized flow,
the large density contrast across the IF, and the constant density through the PDR. The
offset is then easily measured as the distance between the peak Br$\gamma$ and H$_2$ emission. 

{Finally, it is worth noting that the models are 1D whereas the real objects are 3D and so these spatial profiles will not capture the effects of projecting the 3D emission profiles along the line of sight. This will have the effect of blurring out the profiles since sight lines will probe ionized material at different locations in the photoevaporative flow and neutral material. However, since the radii of curvature of the objects we consider are generally $\gtrsim 5$ times the offset size, the effect of projection along the line of sight will be relatively small on the measured offset size. Accurately projecting along the line of sight would require a radiative transfer calculation of the H$_2$ through the pillar which is beyond the scope of the current project. For other successful examples of using 1D models to interpret observed spatial line profiles in photo-evaporative flows, see, e.g., \citet{sankrit, mcleod2016}. However, 1D models might lead to erroneous interpretations of spatially resolved line ratios as discussed in \citet{ercolano2011}.}

\begin{figure}
\includegraphics[width=0.47\textwidth]{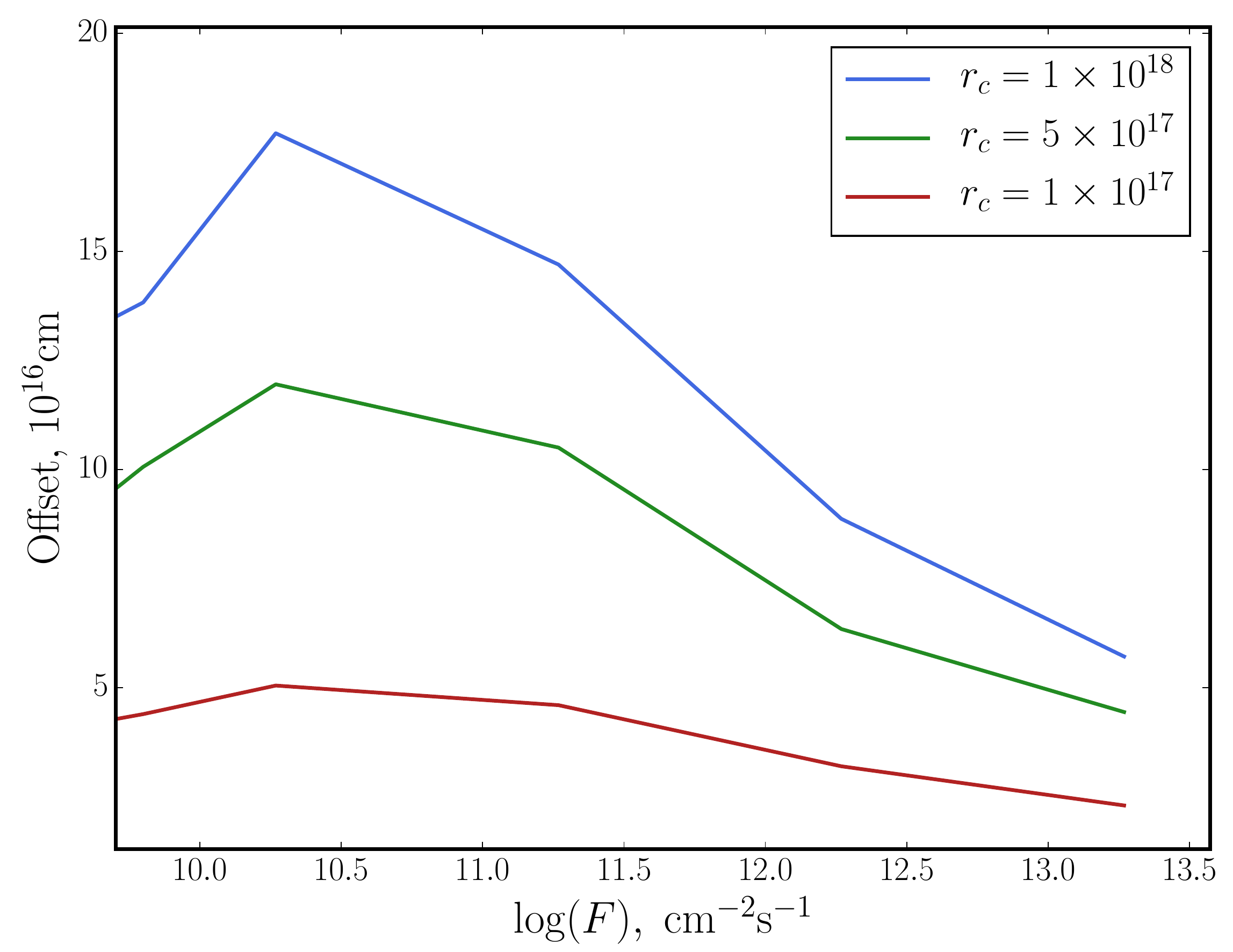}
\caption{Predicted model offsets between Br$\gamma$ and H$_2$
for different radii of curvature and incident fluxes. }
\label{fig:offset_model}
\end{figure}

Figure \ref{fig:offset_model} shows the results of these models for a variety of values of $r_c$ and $F$. The model offsets always lie within roughly a factor of two of
$10^{17}$cm even over a wide range in parameters, in agreement with the observations.
With the exception of the left edge,
the offsets decrease with increasing $F$ and decreasing $r_c$. This trend arises from
the increase in density in the shocked layer that occurs by increasing $F$ or
decreasing $r_c$ (see Figure \ref{fig:model_2}). Increasing the density leads to a thinner PDR
because the thickness of the PDR is set by the column density of hydrogen and
a thinner PDR leads to a smaller offset. The decrease in offsets for log(F) $\lesssim$ 10.4
in Figure \ref{fig:offset_model} is caused by the incident radiation being so weak that a distinct
layer of HI is not formed in the \textsc{Cloudy} models and so the H$_2$ emission
comes from very close to the IF and the offset declines. At these low
incident fluxes, the static models of the PDR are probably inadequate as the advection of
hydrogen through the PDR will start to matter \citep{bertoldi1996}.  It is possible for the ionization
and dissociation fronts to merge under certain conditions \citep[e.g.][]{storzer1998, henney2007},
however our static models are certainly inadequate to describe this situation. 

\section{Conclusions}\label{sec:conclusions}

In this work, we present sky-subtracted large-scale mosaics in H$_{2}$ and Br$\gamma$
for three regions of massive star formation: Cyg~OB2, Carina, and IC~1396. The three
regions span roughly three orders of magnitude in ionizing flux and so provide 
a powerful means to investigate the physics that underlies the
spatial profiles and offsets we observe in the dozens of PDR fronts that populate these regions.

Each of the three star forming regions contains PDRs that have
a wide variety of geometrical shapes,
ranging from more or less spherical globules to highly-elongated cylindrical pillars.
To understand how the shapes of the interfaces affect what is observed in the images, we
fit power laws to the spatial profiles of continuum-corrected Br$\gamma$ flux along 
several hundred lines of sight and grouped the results according to the overall shapes of the
irradiated interfaces. The results confirm model predictions
that the shape of the object strongly influences the spatial profile of the emission in a 
photoevaporative flow. Br$\gamma$ fluxes from cylindrically-shaped
clouds have a relatively flat radial power law index of $-2.54\pm0.16$, 
while spherically-shaped clouds decline more rapidly with distance, with
an average power-law index of $-3.45\pm0.22$. The globule power-law exponents agree well with the expected values between $-3$ and $-4$, although the pillar power-law exponents are somewhat higher than the expected values between $-1$ and $-2$, predicted for photoevaporative flows of constant
velocity. Broadly speaking, the results are consistent with model expectations that more spherical
clouds should exhibit sharper radial decline in the Br$\gamma$ emission
as compared with cylindrical clouds. 

We observe spatial offsets  
between the peaks of the H$_{2}$ and Br$\gamma$ emission throughout 
each of the three star formation regions, always in the sense that
the H$_2$ emission is located deeper into the molecular cloud, underneath
a layer of Br$\gamma$ emission that traces a photoevaporative flow.
By extracting several hundreds of
slices across a wide variety of distinct PDR interfaces, we
found that the spatial offsets between these two lines are
always on the order of $10^{17}$~cm
for all the objects in all three regions.

After estimating the incident ionizing flux, $F$,
received by each object and measuring an
approximate radius of curvature $r_c$ for each object,
we investigated how these two variables correlate with the
observed spatial offsets between Br$\gamma$ and H$_2$.
There is a significant increase in the spatial offset as 
$r_c$ increases. A weak trend of decreasing offset with $F$
is also observed.
Individual geometrical complications introduce significant scatter to these trends.

To investigate the physical processes that determine why the offsets form and what
sets their sizes, a simple theoretical model of Br$\gamma$ emission from a
photoevaporative flow is developed by solving the time-independent Euler and ionization equations.
These density profiles are then used in the numerical code \textsc{Cloudy}
to predict synthetic spatial line profiles for the H$_2$ emission.
The models reproduce all the main results in the data, including the size
of the spatial offsets between H$_2$ and Br$\gamma$, the
positive correlations of the offsets with the radius of curvature of the object,
and even the weak negative correlation of the offset with the amount of incident ionizing
flux.

Overall, these observations support the general picture of molecular
clouds being photoevaporated by young massive stars and lend some confidence that the basic physics that underlies these systems is understood. However, it is clear that
the geometries of the interfaces can be very complex, and this complexity has a significant
effect upon what is observed. Both the H$_2$ and Br$\gamma$ lines will be easily observable 
at high spatial resolution in PDRs with the James Webb Space Telescope, and all of the
analyses done here will be possible to extend to much smaller spatial scales as long as
continuum images are also acquired to enable accurate background subtraction of the
ambient reflected starlight. 

\acknowledgments
This research was supported by Department of Energy's
NNSA division under the NLUF grant DE-FOA-0001109,
and by NASA's National Radio Astronomy student support grant SOSPA3-018.
Financial support for this project was provided by the Department of Energy
grant GR523126.
Based on observations at Cerro Tololo Inter-American Observatory, National Optical Astronomy Observatory (NOAO 11A-0234; PI: P. Hartigan), which is operated by the Association of Universities for Research in Astronomy (AURA) under a cooperative agreement with the National Science Foundation. 
Based on observations at Kitt Peak National Observatory, National Optical Astronomy Observatory (NOAO 12B-0166; PI: P. Hartigan), which is operated by the Association of Universities for Research in Astronomy (AURA) under cooperative agreement with the National Science Foundation. 

{\software{Cloudy \citep{ferland2013}, NEWFIRM Pipeline \citep{swaters2009}}}

\bibliographystyle{aasjournal}
\bibliography{references}

\appendix
\section{List of Objects}
\label{sec:appendix}
The specific interfaces analyzed are listed below along with various calculated parameters.

\begin{deluxetable}{cccccccccc}
\tablecaption{List of interfaces analyzed. $a$ is the best-ftting (negative) log slope of the Br$\gamma$ profile, $r_c$ is the estimated radius of curvature, and $\log(F_{\rm inc})$ is the calculated incident flux of ionizing photons at the location of the interface from the O/B stars in the region. Regions with no entry for $a$ are regions that were not included in the analysis of Section \ref{sec:brgProfiles} because slices longer than the radius of curvature could not be found but were included in the analysis of Section \ref{sec:offsets}. }\label{tab:interfaces}
\rotate
\tablehead{
\colhead{Object} & \colhead{Region} & \colhead{RA} & \colhead{Dec} & \colhead{$a$} & \colhead{$r_c$, $10^{16}$cm} & \colhead{$\log(F_{\rm inc})$, cm$^{-2}$ s$^{-1}$} & \colhead{Average Offset, 10$^{16}$ cm} & \colhead{Number of Slices} & \colhead{Morphological Type} 
}
\startdata
E1 & Carina & 10:46:00 & -59:47:30 & 2.80 & 87.0$\pm8.7$ & 11.96$\pm0.27$ & 12.02$\pm0.84$ & 19 & Pillar \\ 
EC1 & Carina & 10:44:31 & -59:39:50 & 2.60 & 13.8$\pm1.4$ & 12.30$\pm0.28$ & 6.78$\pm0.51$ & 13 & Globule \\
EC2 & Carina & 10:44:33 & -59:34:57 & 3.76 & 16.6$\pm1.7$ & 12.36$\pm0.31$ & 6.62$\pm0.46$ & 26 & Intermediate \\ 
EC3 & Carina & 10:44:57 & -59:37:25 & 3.55 & 17.9$\pm1.8$ & 11.98$\pm0.26$ & 9.24$\pm1.20$ & 7 & Globule \\ 
EC6 & Carina & 10:44:40 & -59:37:52 & 2.12 & 9.7$\pm1.0$ & 12.15$\pm0.20$ & 11.99$\pm1.12$ & 14 & Pillar \\ 
EC7 & Carina & 10:44:51 & -59:40:51 & 3.95 & 19.3$\pm1.9$ & 12.20$\pm0.22$ & 9.76$\pm0.84$ & 11 & Intermediate \\ 
EC8 & Carina & 10:45:08 & -59:39:14 & 3.98 & 19.3$\pm1.9$ & 12.23$\pm0.39$ & 7.71$\pm0.96$ & 10 & Globule \\ 
FW1 & Carina & 10:41:46 & -59:45:14 & -- & 93.8$\pm9.4$ & 11.43$\pm0.16$ & 11.39$\pm1.49$ & 11 & Globule \\ 
N1 & Carina & 10:44:47 & -59:27:22 & 2.19 & 12.42$\pm1.2$ & 11.82$\pm0.21$ & 4.20$\pm0.66$ & 17 & Pillar \\ 
N3 & Carina & 10:45:20 & -59:27:36 & 3.42 & 19.3$\pm1.9$ & 11.89$\pm0.15$ & 4.47$\pm0.33$ & 12 & Globule \\ 
S12 & Carina & 10:43:50 & -59:55:23 & 2.04 & 34.5$\pm3.5$ & 11.94$\pm0.27$ & 12.21$\pm2.39$ & 8 & Pillar \\ 
S10 & Carina & 10:43:53 & -59:57:59 & 2.41 & 46.9$\pm4.7$ & 11.93$\pm0.24$ & 7.20$\pm0.67$ & 13 & Globule \\ 
S1 & Carina & 10:44:40 & -59:57:33 & 2.61 & 30.4$\pm3.0$ & 12.21$\pm0.39$ & 8.11$\pm0.98$ & 19 & Pillar \\ 
S2 & Carina & 10:45:23 & -59:58:19 & 3.25 & 33.1$\pm3.3$ & 11.98$\pm0.43$ & 5.85$\pm0.83$ & 8 & Intermediate \\ 
S6 & Carina & 10:45:59 & -60:05:49 & 2.47 & 37.3$\pm3.7$ & 11.58$\pm0.12$ & 12.89$\pm1.2$ & 19 & Pillar \\ 
W11 & Carina & 10:43:06 & -59:32:13 & 2.34 & 24.8$\pm2.4$ & 12.17$\pm0.19$ & 5.96$\pm0.53$ & 9 & Globule \\ 
W12g & Carina & 10:43:30 & -59:38:13 & 4.25 & 19.3$\pm1.9$ & 12.04$\pm0.21$ & 6.09$\pm1.04$ & 7 & Globule \\ 
W12p & Carina & 10:43:36 & -59:38:39 & 2.52 & 12.4$\pm1.2$ & 12.09$\pm0.22$ & 8.95$\pm2.11$ & 10 & Pillar \\ 
W14 & Carina & 10:43:18 & -59:30:16 & 3.63 & 29.0$\pm2.9$ & 12.15$\pm0.22$ & 4.12$\pm0.30$ & 16 & Globule \\
W2 & Carina & 10:43:33 & -59:34:30 & -- & 115.9$\pm11.6$ & 12.57$\pm0.24$ & 11.87$\pm1.19$ & 16 & Globule \\
W5 & Carina & 10:43:18 & -59:29:40 & 4.41 & 29.0$\pm2.9$ & 12.01$\pm0.25$ & 4.76$\pm0.41$ & 13 & Globule \\
W7 & Carina & 10:44:04 & -59:30:23 & 2.40 & 11.0$\pm1.0$ & 12.67$\pm0.29$ & 5.60$\pm0.97$ & 14 & Intermediate \\
W8 & Carina & 10:44:01 & -59:30:26 & 4.34 & 29.0$\pm2.9$ & 12.62$\pm0.30$ & 4.25$\pm0.39$ & 22 & Globule \\
F1 & Cyg OB2 & 20:35:09 & 41:15:31 & 3.73 & 45.5$\pm4.6$ & 11.92$\pm0.14$ & 8.60$\pm1.13$ & 8 & Globule \\
F2 & Cyg OB2 & 20:30:31 & 41:15:32 & 4.05 & 33.1$\pm3.3$ & 12.21$\pm0.69$ & 6.88$\pm1.15$ & 19 & Globule \\
F4g & Cyg OB2 & 20:34:46 & 41:14:46 & 2.11 & 13.8$\pm1.4$ & 12.00$\pm0.18$ & 6.88$\pm1.23$ & 5 & Globule \\
F4p & Cyg OB2 & 20:34:50 & 41:14:42 & 2.45 & 8.3$\pm0.8$ & 11.07$\pm0.24$ & 5.66$\pm0.66$ & 9 & Pillar \\
F6 & Cyg OB2 & 20:34:15 & 41:08:12 & 3.70 & 9.7$\pm1.0$ & 11.86$\pm0.16$ & 9.81$\pm0.71$ & 8 & Pillar \\
M1 & Cyg OB2 & 20:36:03 & 41:39:41 & -- & 84.2$\pm8.4$ & 11.52$\pm0.13$ & 9.24$\pm0.95$ & 14 & Globule \\
T7 & IC 1396 & 21:34:21 & 57:30:22 & -- & 63.5$\pm6.3$ & 9.92$\pm0.51$ & 10.22$\pm1.07$ & 14 & Globule \\
T2 & IC 1396 & 21:37:04 & 57:30:51 & -- & 155.9$\pm15.6$ & 10.73$\pm0.51$ & 9.78$\pm0.51$ & 46 & Globule \\
TE1 & IC 1396 & 21:33:45 & 57:31:52 & -- & 41.4$\pm4.1$ & 9.82$\pm0.51$ & 10.32$\pm0.77$ & 8 & Globule \\
TE2 & IC 1396 & 21:33:33 & 58:03:26 & -- & 88.3$\pm8.8$ & 9.55$\pm0.50$ & 9.91$\pm2.04$ & 5 & Globule \\
\enddata
\end{deluxetable}

\end{document}